\documentclass[twocolumn]{aastex631}

\usepackage{CJKutf8} 
\usepackage{amsmath}

\submitjournal{ApJS}

\shortauthors{Zaritsky et al.}
\shorttitle{Southern SMUDGes}

\begin{document}

\title{Systematically Measuring Ultra-Diffuse Galaxies (SMUDGes). III.  The Southern SMUDGes Catalog}
  
\correspondingauthor{Dennis Zaritsky}
\email{dennis.zaritsky@gmail.com}

\author[0000-0002-5177-727X]{Dennis Zaritsky}
\affiliation{Steward Observatory and Department of Astronomy, University of Arizona, 933 N. Cherry Ave., Tucson, AZ 85721, USA}

\author[0000-0001-7618-8212]{Richard Donnerstein}
\affiliation{Steward Observatory and Department of Astronomy, University of Arizona, 933 N. Cherry Ave., Tucson, AZ 85721, USA}

\author[0000-0001-8855-3635]{Ananthan Karunakaran}
\affiliation{Instituto de Astrof\'{i}sica de Andaluc\'{i}a (CSIC), Glorieta de la Astronom\'{i}a, 18008 Granada, Spain}

\author[0000-0002-5292-2782]{C. E. Barbosa} 
\affiliation{Universidade de S\~{a}o Paulo, Instituto de Astronomia, Geof\'isica e Ci\^encias Atmosf\'ericas, Departamento de Astronomia, Rua do Mat\~{a}o 1225, S\~{a}o Paulo, SP, 05508-090, Brazil}

\author[0000-0002-4928-4003]{Arjun Dey}
\affiliation{NSF's NOIRLab, 950 N. Cherry Ave., Tucson, AZ 85719, USA}

\author[0000-0002-3767-9681]{Jennifer Kadowaki}
\affiliation{Steward Observatory and Department of Astronomy, University of Arizona, 933 N. Cherry Ave., Tucson, AZ 85721, USA}

\author[0000-0002-0956-7949]{Kristine Spekkens}
\affiliation{Department of Physics, Engineering Physics and Astronomy Queen's University Kingston, ON K7L 3N6, Canada}
\affiliation{Department of Physics and Space Science Royal Military College of Canada P.O. Box 17000, Station Forces Kingston, ON K7K 7B4, Canada}

\author[0000-0002-0123-9246]{Huanian Zhang\begin{CJK*}{UTF8}{gkai}(张华年)\end{CJK*}}
\affiliation{Department of Astronomy, Huazhong University of Science and Technology, Wuhan, Hubei 430074, China;}
\affiliation{Steward Observatory and Department of Astronomy, University of Arizona, 933 N. Cherry Ave., Tucson, AZ 85721, USA}

\begin{abstract}
We present a catalog of 5598 ultra-diffuse galaxy (UDG) candidates with effective radius $r_e > 5.3$\arcsec\  distributed throughout the southern portion of the DESI Legacy Imaging Survey covering $\sim$ 15000 deg$^2$. The catalog is most complete for physically large  ($r_e > 2.5$ kpc) UDGs lying in the redshift range $1800 \lesssim cz/{\rm km\ s}^{-1} \lesssim 7000$, where the lower bound is defined by where incompleteness becomes significant for large objects on the sky and the upper bound by our minimum angular size selection criterion. Because physical size is integral to the definition of a UDG, we develop a method {of distance estimation} using existing redshift surveys. With three different galaxy samples, two of which contain UDGs with spectroscopic redshifts, we estimate that the method has a redshift accuracy of $\sim$ 75\% when the method converges, although larger, more representative spectroscopic UDG samples are needed to fully understand the behavior of the method. We are able to estimate distances for 1079 of our UDG candidates (19\%).
Finally, to illustrate uses of the catalog, we present distance independent and   dependent results. In the latter category we establish that the red sequence of UDGs lies on the extrapolation of the red sequence relation for bright ellipticals and that the environment-color relation is at least qualitatively similar to that of high surface brightness galaxies. Both of these results challenge some of the models proposed for UDG evolution.

\end{abstract}


\section{Introduction}
\label{sec:intro}
This is the third paper in a series presenting results from our ongoing search for low surface brightness, physically large galaxies. The previous papers, \citet[][hereafter, Paper I]{Zaritsky+2019} and \citet[][hereafter Paper II]{Zaritsky+2021}, presented both the scientific motivation and description of our methodology. The principal difference between the earlier papers and the current one is that we have progressed beyond developing and demonstrating how we identify and measure these galaxies and now produce a large catalog. We present 5598 candidate ultra-diffuse galaxies (hereafter UDG candidates, with criteria of $\mu_{0,g}>24$ mag arcsec$^{-2}$ and $r_e > 5.3$ arcsec) covering the southern portion of the 
Dark Energy Spectroscopic Instrument (DESI) Legacy Imaging Surveys \citep[hereafter referred to as the Legacy Survey;][]{Dey+2019}, defined as the portion of the Legacy Survey that uses DECam \citep{decam} images obtained with the Blanco 4m telescope. 
We refer to these sources  as UDG candidates because we do not yet have the distance measurements needed to determine their physical size. However, we describe below a method by which we do obtain distance {\it estimates} for 1079 of the candidates and determine that 514 of those do indeed have $r_e > 1.5$ kpc, thereby confirming their nature as UDGs. Further arguments suggest that an even greater majority of the candidates presented in the catalog are likely to be UDGs. We refer the reader to Papers I and II for a description of our scientific interest in UDGs and historical context, but stress here that the observational definition of UDGs is merely a way of selecting objects  that lie on the tails of physical parameter distributions, rather than defining a physically distinct class of galaxy. Nevertheless, such extreme objects have the potential to challenge galaxy evolution models that are tuned to reproduce more typical galaxies.

To illustrate the science potential of this new catalog, we revisit a few of the results described in Paper II based on the much smaller sample available then and we also present results using our new UDG subsample with estimated redshifts. Detailed and complete investigations of the aspects explored here will be provided in future papers and, we hope, by other investigators who exploit this catalog and the upcoming northern extension. We describe the data and reprise the outline of our processing methodology in \S\ref{sec:data} and \S\ref{sec:processing}. Some additional details and the catalog are presented in \S\ref{sec:catalog}. We describe our method for distance estimation in \S\ref{sec:distances} and a few preliminary results in \S\ref{sec:results}. We use the concordance $\Lambda$CDM cosmology when converting to physical units \citep[WMAP9;][]{wmap9} and magnitudes are on the AB system \citep{oke1,oke2}.

\section{The Data}
\begin{figure}[ht]
\begin{center}
\includegraphics[width=0.40\textwidth]{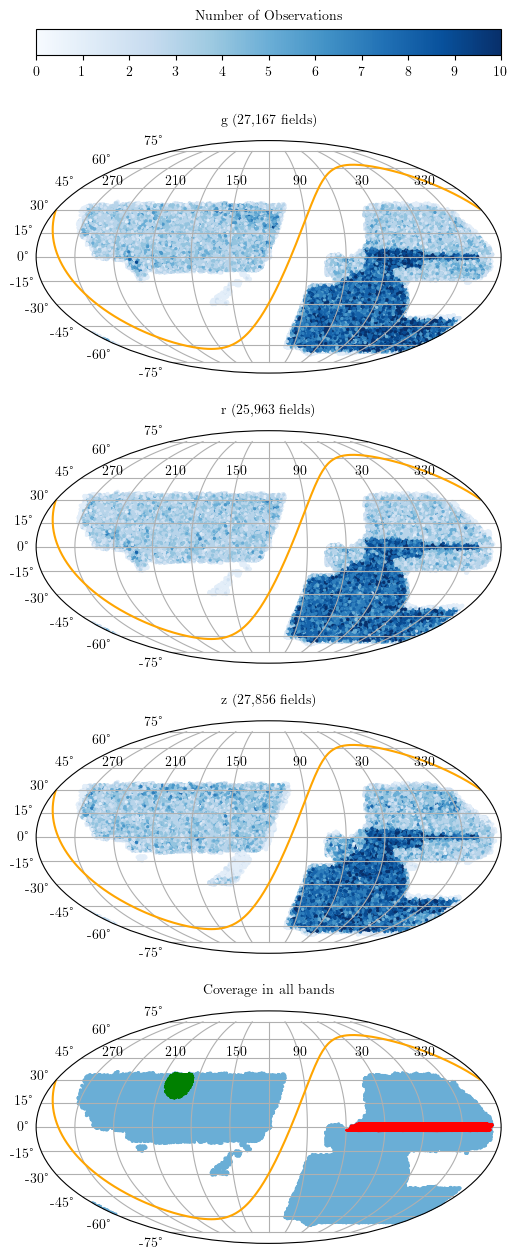}
\end{center}
\vskip -.5cm
\caption{Footprint of the Legacy Survey DR9 observations used in this study in Right Ascension and Declination. In the top three panels, shading denotes the observation density for each band as shown in the top color bar. The bottom panel shows regions with coverage available in all three filters. Footprints for our previous work in Coma (Paper I) and SDSS Stripe 82 (Paper II) are shown in green and red, respectively.  The Galactic plane is traced by the orange curve.}
\label{fig:footprint}
\end{figure}

We report the results of our analysis of all images obtained with DECam \citep{decam} at the CTIO 4m, or Blanco telescope, that are included in Data Release 9 (DR9) of the DESI Legacy Imaging Surveys \citep{Dey+2019}. Briefly, this survey was initiated  to provide targets for the DESI survey \citep[][]{DESI}.  In addition to the images from DECam (referred to as DECaLS), the complete Legacy Survey incorporates observations obtained with an upgraded MOSAIC camera \citep{mzls-camera} at the KPNO 4m, or Mayall telescope, (MzLS, Mayall z-band Legacy Survey)  and the 90Prime camera \citep{90prime} at the Steward Observatory 2.3m, or Bok, telescope \citep[BASS, Beijing-Arizona Sky Survey;][]{bass} to provide deep three-band ($g=24.7$, $r=23.9$, and $z=23.0$ AB mag, 5$\sigma$ point-source limits) images. The Legacy Survey encompasses about 
14,000 deg$^2$ of sky visible from the northern hemisphere between declinations  approximately bounded by $-18^\circ$ and +84$^\circ$. The DECaLS covers about 9,000 deg$^2$ and provides the vast majority of observations obtained between $-18^\circ$ and +32$^\circ$.  DR9 is augmented with deep DES \citep{des} DECam observations\footnote{\href{https://www.legacysurvey.org/dr9/description/}{legacysurvey.org/dr9/description/}} obtained to southern declinations of $-68^\circ$ (Figure \ref{fig:footprint}) of an additional $\sim$ 6000 deg$^{2}$.

Here we present an analysis of only the DECam data. Extending SMUDGes to the full Legacy Survey footprint will require some adjustment and re-certification of the pipeline for MzLS and BASS data. We will present that portion of our catalog in the next data release paper.

\label{sec:data}
\section{Processing}
\label{sec:processing}
All processing and analyses are performed on  the Puma
cluster at the University of Arizona High Performance Computing center\footnote{\href{https://public.confluence.arizona.edu/display/UAHPC/Resources}{public.confluence.arizona.edu/display/UAHPC/Resources}}. The compute nodes on this machine currently contain 94 usable CPUs, 512 GB of RAM, and 1640 GB of solid state storage with half of the storage guaranteed to be available during processing.  All files and observational data used in this study are publicly available at the Legacy Survey\footnote{\href{https://www.legacysurvey.org/dr9/files/}{legacysurvey.org/dr9/files/}} or the NSF’s NOIRLab\footnote{\href{https://astroarchive.noirlab.edu/}{astroarchive.noirlab.edu}} website.  We limit observations to those contained in the file survey-ccds-decam-dr9.kd.fits.gz  which is included in the Legacy Survey's DR9.  This file has information  for  each  CCD  image  used  in  the  data release and excludes those considered inadequate for further processing.  We also make use of the magnitude zero points and image full widths at half maximum (FWHMs) contained in this file which are generated for each CCD by the Legacy Survey's pipeline. The footprint (Figure \ref{fig:footprint}) contains 80,986 observations with acquisition dates ranging from 31 August 2013 to 7 March 2019 and includes 4,991,222 individual CCD images considered adequate for processing.  In compressed form these observations require more than 25 TB of storage which far exceeds the capacity of the processing nodes. Because transferring files between the processing nodes and main storage drives is inefficient, we limit the number of observations that are simultaneously processed to 1,000, which allows all intermediate files needed for processing to be kept on the processing node.  This is accomplished by dividing the observation footprint into individual tiles with sizes determined by the number of included observations rather than area.  Observation centers extend from about $-67.4^\circ$ to  34.8$^\circ$ in Declination and we divide this into 10 equal stripes.  We overlap adjacent stripes by 1.2$^\circ$ in Declination to account for the 2.2$^\circ$ DECam field of view.  Each stripe is then divided into individual tiles in Right Ascension with an objective of maximizing the number of observations within our imposed limit of 1,000.  As with Declination, tiles overlap 1.2$^\circ$ in Right Ascension to account for the field of view.  This process results in 107 tiles which are individually processed.

Other than adaptations made because of the much larger footprint, our processing pipeline is essentially unchanged from that described in Paper II.  The major steps involved in creating our catalog include: 1) image processing to create a list of potential UDGs; 2) screening for cirrus contamination which can create false positives; 3) automated classification of remaining candidates; 4) modeling completeness, biases, and uncertainties using simulated sources; and 5) creating the catalog.  Each of these steps and prior modifications are described in detail in Paper I and Paper II and here we only briefly summarize them.  We also describe further modifications that were implemented because of the expanded footprint.  Unless otherwise stated, total numbers provided below and in Table \ref{tab:screening} include all tiles. Duplicate UDG candidates from overlapping regions are removed prior to our machine classification of the candidates.

\begin{deluxetable*}{lrrr}
\tablecaption{Number of detections and UDG candidates after each processing step}
\label{tab:screening}
\tablewidth{0pt}
\tablehead{
\colhead{Process}&Description in Text&
\colhead{Detections} &
\colhead{UDG Candidates}\\
}
\startdata
Wavelet screening & \S3.1, Step 4 & 453,373,301  & NA \\
Object matching &\S3.1, Step 5  & 267,952,256 & 82,439,516\\
S\'ersic screening &\S3.1, Step 6 & 11,452,256 & 1,853,319\\
Required observations & \S3.1, Step 7& NA & 752,798\\
Initial GALFIT screening & \S3.1, Step 8& NA & 191,933 \\
Final GALFIT screening & \S3.1, Step 9& NA & 67,902  \\
Cirrus screening & \S3.2 & NA & 13,873  \\
Duplicate removal  & \S3.3& NA & 10,892\\
Automated classification & \S3.4& NA & 5,760\\
Visual Confirmation & \S3.5& NA & 5,598\\
\enddata
\end{deluxetable*}

\subsection{Image processing }
Our image processing pipeline for identifying potential UDG candidates consists of the following steps.  Rationales for each of these steps are provided in Paper II and are not repeated here.
\begin{enumerate}
\item We obtain calibrated images and data quality masks that have been processed with the DECam Community Pipeline \citep{valdes}  from the NOIRLab website. Because of tile overlaps, a total of 99,316 observations consisting of 5,897,921 individual CCD images are reprocessed by our pipeline.
\item We use the data quality masks to assist in identifying and removing CCD artifacts and  wings of saturated stars.
\item We model and subtract sources on CCDs that are 2 mag arcsec$^{-2}$ brighter than a specified threshold in each band (24.0 for $g$, 23.6 for $r$, and 23.0 for $z$), thereby removing objects that are clearly too bright to qualify as UDG candidates.
\item We isolate candidates of different angular scales using wavelet transforms with tailored filters.  This results in a total of 453,373,301 detections, or an average of $\sim$77/CCD, the vast majority of which will not be classified as UDG candidates after further screening.
\item We only retain candidates with at least two coincident detections (defined as lying within 2\arcsec\ of their mean centroid) among different exposures, regardless of which wavelength filter was used in the detection to limit spurious detections.  Each group created in this manner is considered to be a unique candidate located at the mean centroid position.  This requirement rejects all but 267,952,256 wavelet detections which comprise 82,439,516 groups.

\item  To limit the number of detections requiring time-consuming coaddition and GALFIT \citep{peng} modeling, we use the LEASTSQ function from the Python SciPy library \citep{jones} to obtain much faster, rough parameter estimates.  We fit an exponential S\'ersic model ($n$=1) to each candidate on a CCD and require that they meet parameter thresholds of $r_e > 4\arcsec$  and $\mu_0$ values of greater than 23.0, 22.0 and 21.5 mag arcsec$^{-2}$ for $g$, $r$, and $z$, respectively. 
These criteria are relaxed relative to those required after our final GALFIT modeling (Step 9) to avoid inadvertently rejecting valid candidates. A total of 11,452,256 detections representing 8,003,489 distinct groups meet these criteria.  However, the majority of groups have only a single surviving member with only 1,853,319 having more than one.

\item For further verification, we require that candidates pass the preliminary screening described in the prior step on at least 20\% of the available observations or a minimum of two observations for those with less than ten.  A total of 752,798 meet this threshold.
\item We perform an initial GALFIT screening of stacked cutouts using a fixed S\'ersic index of $n$ = 1, without incorporating the point spread function (PSF) into the model.  We again use generous thresholds compared to those of Step 9 and accept candidates with $r_e > 4\arcsec$  and $\mu_{0,g} >$ 22.95 mag arcsec$^{-2}$ or $\mu_{0,z} >$ 21.95 mag arcsec$^{-2}$ if there is no available measurement of $\mu_{0,g}$.  A total of 191,933 candidates survive this stage.
\item  We obtain our final GALFIT results using a variable S\'ersic index with an estimate of the PSF incorporated into the model.  Estimates of morphological parameters ($r_e$, $b/a$, $n$, and $\theta$) are derived from a stacked image using all three filters. These morphological parameters are then held fixed when estimating photometric properties.  Because our machine learning classifier (Section \ref{subsec:classifier}) uses information from all three filters, we require that a candidate have at least one observation in each filter. Images from a band are used even if the object was not initially detected on it. The 67,902 candidates that pass our final criteria of $r_e \ge 5.3\arcsec$, $\mu_{0,g}\ge$ 24 mag arcsec$^{-2}$ (or $\mu_{0,z}\ge$ 23 mag arcsec$^{-2}$ if GALFIT failed to model $g$), $b/a \ge $ 0.37, and $n < 2$ form the population used for further screening as described below.

\end{enumerate}
\subsection{Screening of Spurious Sources Caused by Cirrus}
\label{subsec:cirrus}
\begin{figure}[ht]
\begin{center}
\includegraphics[width=0.45\textwidth]{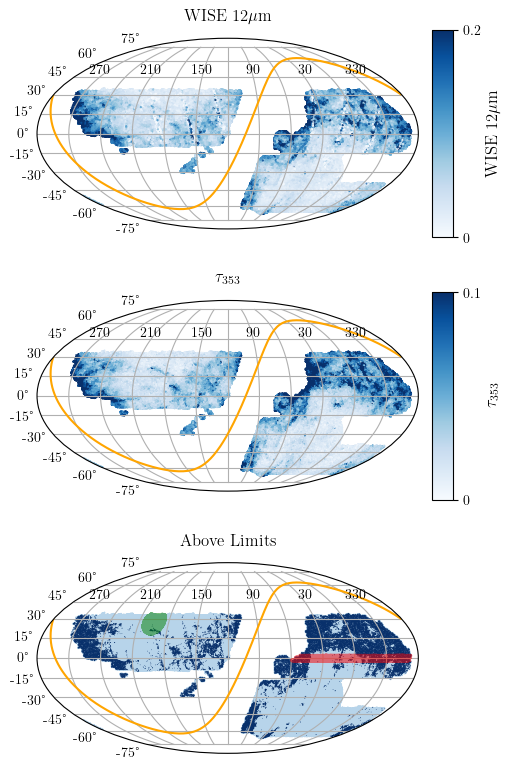}
\end{center}
\vskip -.5cm
\caption{Cirrus contamination within the DECaLS footprint. The top panel shows the distribution of WISE 12 $\mu$m while the middle panels show it for $\tau_{353}$. Regions in dark blue in the bottom panel exceed our dust proxy thresholds of either 0.1 MJy/sr for WISE 12 $\mu$m or 0.05 for  $\tau_{353}$ and comprise $\sim$30\% of the entire footprint. Overlays in green and red in the bottom panel correspond to the footprints from our previous work in Coma (Paper I) and SDSS Stripe 82 (Paper II), respectively.  The Galactic plane is shown in orange.}
\label{fig:dust}
\end{figure}

Our work on Stripe 82 (Paper II) showed that large regions of the survey footprint are contaminated by Galactic cirrus that can result in detections that are sometimes difficult to differentiate from legitimate UDG candidates. To address this challenge, we developed a screening process for probable cirrus contamination that makes use of $\tau_{353}$ \citep{planck} and WISE 12 $\mu$m \citep{meisner} dust maps.  In particular, we extract single point values from each dust map located at the coordinates of a candidate and reject those with values exceeding 0.05 for $\tau_{353}$ or 0.1 MJy/sr for WISE 12 $\mu$m.  With 43.4\% of the Stripe 82 footprint from Paper II exceeding these thresholds, dust contamination was a major factor in determining completeness in that region.  In contrast, only 1.7\% of the Coma region studied in Paper I exceeds these thresholds, demonstrating the large variations found within the DECaLS footprint.  This conclusion is visually confirmed in Figure \ref{fig:dust}, which shows that $\sim$31\% of the entire DECaLS footprint exceeds our thresholds, primarily in regions immediately adjacent to the Galactic plane.  Using our criteria, we reject 54,029 candidates ($\sim$80\%) that survived our image processing pipeline as potential false detections caused by dust, leaving 13,873. The fraction of candidates rejected far exceeds the fraction of the DECaLS footprint that fails our dust criteria, emphasizing the problem with false positive detections caused by cirrus contamination.

\subsection{Screening of Duplicates}
\label{subsec:duplicates}

Before classifying the remaining candidates, we eliminate duplicate entries that result primarily from our defined overlapping tiles.  We consider any candidates lying within 10\arcsec\ of each other to to be duplicate detections.  While this theoretically could cause the loss of some closely spaced UDGs, we visually inspected all 75 cases of duplicates lying between 5\arcsec\ and 10\arcsec\ and found none containing {\sl bona fide} separate candidates. The rejected systems  consisted either of cirrus that was not rejected by our dust criteria, tidal material, or very large candidates that the processing broke up and identified as separate sources. We discuss our incompleteness in very large sources in \S\ref{completeness}. 

We follow a two step protocol for handling duplicates.  We select whichever of the multiple sources has the greater number of observations and reject the remainder.  If two or more have the maximum number of observations, we create a new entry using the median values of all parameters. Finally, because our machine learning classifier uses information from all three filters, we require that a candidate have at least one observation in each of filter, leaving us with 10,892 sources.

\subsection{Automated Classification}
\label{subsec:classifier}
Details on our approach for computer classification are described in detail in the appendix of Paper I with modifications addressed in Paper II.  Briefly, we found that our best results are obtained with the TensorFlow Keras version of the convolutional neural network, EfficientNetB1 \citep{efficientnet} trained on 224 $\times$ 224 pixel ($\sim 59\arcsec \times 59\arcsec$) cutouts downloaded from the Legacy Survey.  Using this protocol we achieved an overall accuracy of 96.2\% (513/533) on a test set with 8 false positives (specificity of 96.5\%) and 12 false negative classifications (sensitivity of 96.1\%). As described in Paper II, training and test sets were derived from both our Coma and Stripe 82 data with depth distributions that should approximate those of the current footprint shown in Figure \ref{fig:footprint}.  Therefore, we make no changes for the current study and use the prior trained network resulting in 5,860 candidates classified as potential UDGs.  As mentioned in Paper II we find occasional sources structurally similar to other candidates but significantly redder than the Coma cluster red sequence. Although some of these may be objects of interest in their own right (e.g., high redshift Ly $\alpha$ nebulae), we conclude based on visual inspection that these are unlikely to be UDGs and so reject those with $g-r$ colors $>$1.0 mag.  A total of 100 candidates fail to meet this threshold, leaving 5,760 catalog entries. Applying the completeness corrections described further below, this number of cataloged candidates corresponds to 15830 candidates covering the $\sim$ 15000 deg$^{2}$ survey footprint, for a surface density of $\sim$ 1 candidate per deg$^{2}$.

\subsection{Visual Confirmation}
\label{subsec:confirm}
In Paper II we concluded from visual examination that about 2.6\% (8/306) of the candidates identified as potential UDGs by our automated classifier in a test set were false positives.  To minimize the effects of false positives in our current catalog, objects classified as candidate UDGs were visually reviewed by two authors (DZ and RD) in three steps.  To estimate consistency, both reviewers initially classified the same random 10\% (576) of the sample.  Of these, 10 (1.7\%) are classified as false positives by both reviewers with disagreements on another 10.  Because UDGs may randomly fall in close proximity to an unrelated normal galaxy along the line of sight, we find that most disagreements involve differentiating candidates from tidal material or spiral arms.  Although without additional information (distance measurements, higher resolution images, etc.) it is sometimes impossible to be certain, we attempt to add some objectivity to this decision: for a candidate to be assigned to another galaxy, we require that it have either a clear bridge to the purported parent or appear to be part of a shell surrounding the parent.  We then split the remaining candidates into 2 groups of 2,592 each which either RD or DZ visually evaluated.  This step is intended to catch those with obviously incorrect designation. Of the 5,184 candidates inspected by either RD or DZ, a total of 79 (1.5\%) were classified as false positives and 231 (4.5\%) were flagged for further evaluation.  The false positives include two large candidates with duplicate detections whose centers are separated by greater than the 10\arcsec distance required for our duplicate screening (\S\ref{subsec:duplicates}). The final step consisted of evaluating these 231, together with the 10 disagreements from the first set, in more detail by both reviewers. A total of 57 are labeled as false positives by both reviewers and disagreements remain on 16.  Because we want to minimize the number of false or ambiguous detections in our catalog, we consider any disagreements to be false positives resulting in 162/5,760 (2.8\%) being labeled as such.  This fraction is nearly identical to that presented in Paper II suggesting that our earlier training and test sets are appropriate for the current data.

\subsection{Estimating completeness, biases, and uncertainties using simulated UDGs}
In Paper II we planted simulated UDGs at random locations to estimate uncertainties and recovery completeness.  Simulated sources were placed at an average surface density of 2000 deg$^{-2}$ (about 100 per CCD) using S\'ersic profiles with random structural and photometric properties. We process the sources separately with the same pipeline as for our real sources, including the automated classification.  We estimated completeness and both structural and photometric uncertainties using polynomial models created with the PolynomialFeatures function from the Python Scikit-learn library \citep{sklearn} and a four layer neural network implemented with Keras \citep{keras}.  Details about our protocol for developing the models and selection of their parameters are given in Paper II and, other than modifications and information needed for understanding results, will not be repeated here.

We initially used uniform distributions for all parameters but found that the number of faint simulations surviving our pipeline was inadequate for robust statistics and, therefore, augmented the initial run with two more using normal distributions with a mean of $\mu_{0,g} = 26.4$ mag arcsec$^{-2}$ and a standard deviation of $\sigma = 1.3$ mag arcsec$^{-2}$.  We now incorporate these into a single run consisting of 1/3 with a flat distribution and 2/3 with normal distributions.  Because the full DECaLS footprint is so much larger than that of Stripe 82, we decrease the simulation density to 600 deg$^{-2}$ or about 30 per CCD.  In another departure from Paper II, we now account for tile overlaps by assigning both simulated sources and pipeline survivors to a single tile with borders defined by the midpoints of the overlapping regions.  Points lying outside of this region are ignored.  This is necessary because tile size is determined by a maximum number of observations and regions with higher observation densities have smaller tiles with a relatively larger fraction overlapping, resulting in a bias towards such regions. Furthermore, because we require our science candidates to be imaged at least once in all three filters, we only include the 7,090,079 simulations meeting this same criterion (see Section \ref{subsec:duplicates}). As described in Paper II, in order to obtain adequate statistics our methodology prevents us from modeling the full range of our simulations.  Because we want our models to include the expected ranges of our science candidates, we use  expanded criteria for simulations ($23.5 <\mu_{0,g}< 27.5$ mag arcsec$^{-2}$, $3.5 <r_e < 20$\arcsec, $b/a > 0.25$, and $0.1 <n < 2$) with 1,142,617 meeting these thresholds.

As with our real candidates, we use cutouts downloaded from the Legacy Survey and centered on the detection when applying our automated classification.  Because real sources may occasionally be incorrectly associated with a simulated one, we make two passes through our automated classification network. We initially evaluate cutouts before placing our simulations on them. Any that pass our classifier at this step cannot be simulations and are rejected from further consideration. We then place the simulations on these same cutouts and reclassify them. Using these criteria, a total of 956,604 of the original simulations are classified as candidates and are used for estimating uncertainties and completeness.  

As noted in Paper II, our ability to estimate uncertainties and completeness is limited by the differences between our smooth S\'ersic models and real UDGs, some of which may have complex morphologies.  Nonetheless, they do provide baseline estimates of these parameters which may be augmented in the future by comparing our results to other surveys using different observational and analytic strategies. 

\subsubsection{Uncertainties}
We define the parameter error for a given simulated source as the difference between the final GALFIT value and the value used when creating the simulated source (GALFIT $-$ input). Errors are generally asymmetric and we define the bias as the median difference and the ``1$\sigma$" confidence limits as the 15.1 and 84.9 percentiles of the distribution for a set of similar simulated objects. 

Polynomial models, especially of high order, may extrapolate very poorly for data points lying outside of the fitted range. Other than for $r_e$, which has an upper limit of 20\arcsec, our simulation range for individual parameters extends beyond those expected for our science targets.  However, the models do not include all possible combinations in the 4-D parameter space, which is limited by the number of simulations surviving our entire pipeline.  Therefore, candidates with parameters that are within individual simulation limits may still fall out of the full parameter space when combined.   Our models for Stripe 82 used a 9\textsuperscript{th} degree polynomial which, except for a few candidates, gave us acceptable uncertainty estimates for parameters that fell out of the modeling range.  Because of the much larger area of observation and a more diverse population, the range of GALFIT estimates for all parameters in the current study exceed those found in Stripe 82. This problem is particularly acute for $r_e$ where GALFIT estimates may produce values several factors greater than our simulation maximum of 20\arcsec. We find that using a  9\textsuperscript{th} degree polynomial results in meaningless uncertainty estimates for a significant number of candidates with parameters lying outside of the fitted model.  Because we use a much larger number of simulations in the current study, even low order polynomials fit our model data points better than the 9\textsuperscript{th} degree polynomial did for Stripe 82 and we select a 2\textsuperscript{nd} order polynomial for this study.  As discussed in Paper II, there is negligible bias in our $\theta$ determinations and to avoid adding noise, we set all $\theta$ biases to zero in the catalog.

\subsubsection{Completeness}
\label{completeness}
Completeness is defined as the probability that a candidate with given structural and photometric parameters will be identified as such after passing through our entire pipeline, and is assessed using four modeled parameters ($\mu_{0,g}$, $r_e$, $b/a$, and $n$).  We again use a 2\textsuperscript{nd} degree polynomial to fit the simulation results rather than the 3\textsuperscript{rd} degree used for Stripe 82. Bias corrections are applied to our catalog entries before estimating their completeness probabilities.  

A limitation of our completeness analysis arises from the training sample used for automated classification.  This was drawn from candidates in the Coma and Stripe 82 regions, which contained very few visually confirmed candidates with angular extents $>$30\arcsec\ and, therefore, the vast majority of ``large" candidates were classified as artifacts or tidal material.  To estimate the effect of this limitation, we visually examined all 57 candidates with GALFIT $r_e >$25$\arcsec$ in the tile containing the Virgo cluster prior to our automated classification. The Virgo cluster is only about one-sixth the distance to the Coma cluster and a candidate with our target minimum physical extent of 2.5 kpc at Coma would have an angular extent of $\sim$32$\arcsec$ in Virgo. Therefore, all the galaxies that we would classify as UDGs in the Coma cluster would be susceptible to misclassification in the Virgo cluster. Of the 23/57 that we  visually determine to be legitimate candidates, none (0/10) of those with $r_e >$33$\arcsec$ were correctly classified by our automated classifier. The recovery rate improves to 50\% (3/6) for candidates with  $30\arcsec < r_e < 33\arcsec$ and to 71\% (5/7) for those with $25\arcsec < r_e < 30\arcsec$. 

This incompleteness is unfortunate, but not highly problematic for the survey as a whole. The principal goal of SMUDGes is to explore the nature of large ($r_e >$ 2.5 kpc) UDGs across environments. Accurate distances are required to determine physical parameters, and those galaxies that have low recessional velocities are plagued by large distance uncertainties due to unknown peculiar velocities. Our calculated completeness estimates extend to $r_e = 20$\arcsec. This angular size corresponds to 2.5 kpc at $z \sim 0.006$ or $cz = 1800$ km sec$^{-1}$. This lower bound excludes Virgo ($cz \sim$ 1100 km sec$^{-1}$), and even so is still somewhat of a low recessional velocity for an accurate (i.e., likely to be good to 10 to 20\%) Hubble flow distance determination. As such, we conclude that our catalog best reflects the population of large ($r_e > 2.5$ kpc) UDGs beyond $cz \sim 1800$ km sec$^{-1}$. The upper bound on $cz$ over which we expect to  find large UDGs simply comes from determining when large UDGs begin to fall below our 5.3 arcsec angular size criterion, which occurs at $cz \sim$ 7000 km sec$^{-1}$. Nevertheless, the catalog  also contains smaller UDGs, with $r_e$ between 1.5 and 2.5 kpc, and UDGs beyond $cz \sim 7000$ km sec$^{-1}$.

\section{The Catalog}
\label{sec:catalog}

To keep our simulation pipeline identical to our science pipeline, we include all 5,760 candidates in our catalog with those visually identified as false positives flagged.  Although a relatively small fraction, these should be omitted from any conclusions drawn from our results that do not depend sensitively on the completeness fractions.  Descriptions of the catalog entries are presented in Table \ref{tab:catalog} with the full catalog available in the electronic version of the Table. The parameter entries include their GALFIT estimates as well as their bias and confidence limits produced by our models.  We additionally flag any entry where we had to extrapolate the fitted model beyond the range of the constraints.  Even when using lower order polynomials, we find that uncertainty and completeness estimates may be unacceptable when a candidate has an estimated $r_e > 30\arcsec$ and, therefore, we flag these by setting these estimates to $-$99.9999 for candidates with $r_e$ values exceeding this limit. Users are encouraged to apply the bias values by subtracting those presented in the catalog from the corresponding uncorrected measurements when drawing conclusions from the data. However, uncertainties and completeness estimates for flagged entries are suspect and should be used with caution (\S3.6). Note that this limitation results in severe incompleteness in physically large UDGs in nearby clusters such as Virgo.

\begin{figure*}
    \centering
    \includegraphics[scale=0.45]{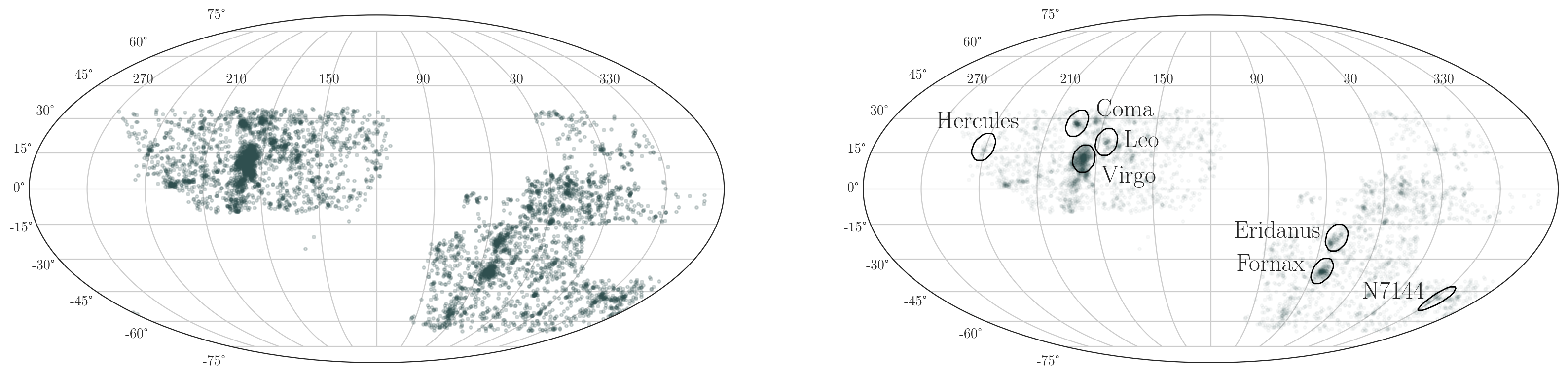}
    \caption{\textit{The current SMUDGes sample of 5598 UDG candidates on the sky in equatorial coordinates. Top panel shows the individual candidates more clearly, while the lower highlights overdensities. Several well-known structures are circled and labeled, with the strongest concentration being the Virgo cluster, even though the catalog is significantly incomplete in large UDGs for low redshifts ($cz < 1800$ km sec$^{-1}$) because they are so large on the sky. The size of the projected circles have no physical meaning.}}
    \label{fig:skyplot}
\end{figure*}

Parameters are corrected for bias before their completeness values are estimated. Completeness estimates may be suspect (completeness flag $\ne 0$) for either of two reasons.  The parameters may be outside of the parameter space defined by our completeness model and these have flag = 1.  Alternatively, the bias correction derived from the uncertainty model may be unreliable and these have flag = 2.  In either case, the results should be used with caution.

Photometric parameters are not corrected for extinction, but extinction values are included in the table for those who wish to use them. Our extinction estimates (A$_g$, A$_r$, A$_z$) are calculated using the SDSS $g$, $r$, and $z$ Legacy Survey extinction coefficients\footnote{\href{https://www.legacysurvey.org/dr9/catalogs/\#galactic-extinction-coefficients}{legacysurvey.org/dr9/catalogs/\#galactic-extinction-coefficients}} which differ slightly from those in Table 6 of \cite{Schlafly}. E(B-V)$_{SFD}$ is estimated using the dustmaps.py \citep{green} SFD dust map based on the work of \cite{SFD}. 

Finally, we recognize that even in situations where both of our reviewers visually classify a candidate as a potential UDG, in a population of this size some are going to be ambiguous and other observers may think otherwise.  This number should be small and not significantly affect any conclusions drawn from analyses of the entire unflagged data set.  However, we obviously would recommend that images be reviewed for any studies drawing conclusions based on individual candidates, particularly if those are extreme in any way (e.g., largest, faintest, etc.).

\subsection{Comparison to Previous Catalogs}
\label{sec:catalog_comp}

In Paper II we presented a comparison to the \cite{Greco+2018a} and \cite{Tanoglidis+2021} catalogs over the Stripe 82 region. Here, with our newly realized greater areal coverage, we expand that comparison and also now include the \cite{Prole+2019} catalog.
As we stressed in Paper II, catalogs tend to have different strengths and weaknesses so the comparison here is not meant to place any catalog above another, but rather to assess the robustness of the catalogs and highlight where the potential advantages of using the SMUDGes catalog might lie.

\begin{figure}
    \centering
    \includegraphics[scale=0.4]{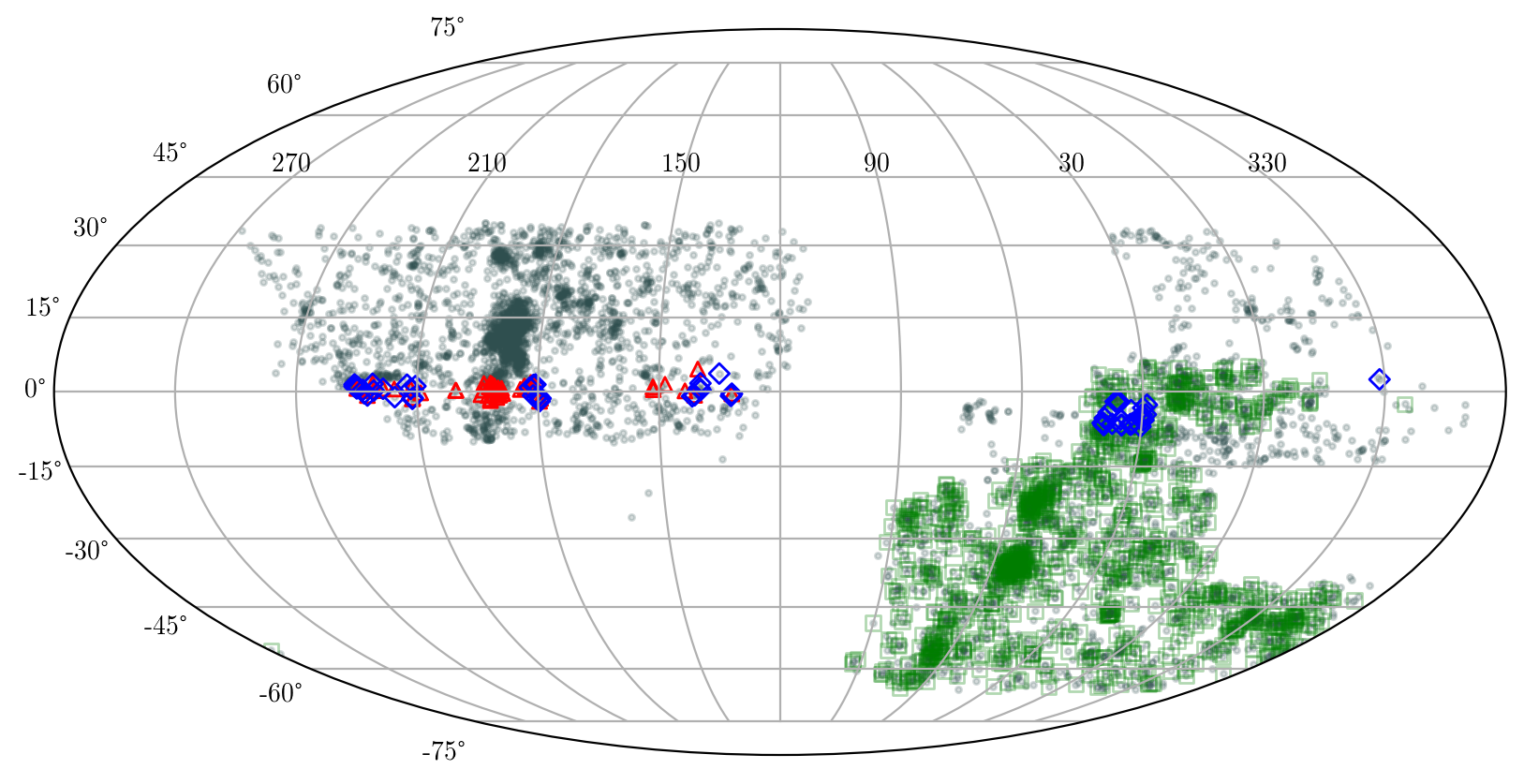}
    \caption{\textit{Distribution of matched UDG candidates on the sky for three different catalogs compared to the current SMUDGes catalog distribution. The blue diamonds correspond to galaxies from \cite{Greco+2018a}, the red triangles from \cite{Prole+2019}, and the green squares from \cite{Tanoglidis+2021}. The small dots represent the SMUDGes candidates.}}
    \label{fig:catalog_comp}
\end{figure}

In Figure \ref{fig:catalog_comp} we present the distribution of matched sources from the three catalogs and SMUDGes on the sky. The first clear difference is that SMUDGes covers a much larger area of sky than the \cite{Greco+2018a} and \cite{Prole+2019}  catalogs and provides more northern coverage than the \cite{Tanoglidis+2021} catalog. 
In detail the various catalogs differ from SMUDGes in other ways as well. The \cite{Greco+2018a} catalog includes many more systems of small angular extent (only 23\% of the galaxies satisfy the SMUDGes $r_e > 5.3$\arcsec\ criterion) and a number of higher central surface brightness galaxies (only 67\% satisfy the the SMUDGes $\mu_{0,g} > 24$ criterion, assuming the global $g-i$ color is representative of the central color). On the other hand, the \cite{Greco+2018a} catalog has significantly better representation at the faintest central surface brightnesses where SMUDGes begins to become significantly incomplete.
As such, the two samples, even within the overlapping regions of sky, are nearly distinct with only 62 sources in common. Similarly, the \cite{Prole+2019} sample also is dominated by source of smaller angular extent than the SMUDGes criterion and only 57 objects are matched to the SMUDGes catalog. Of the 66 sources in that catalog with $r_e > 5.3$\arcsec, 57 are matched, demonstrating that the mutual completeness is high when comparing similar populations and that the low number truly reflects the differences in the selection.

\begin{figure}
    \centering
    \includegraphics[scale=0.4]{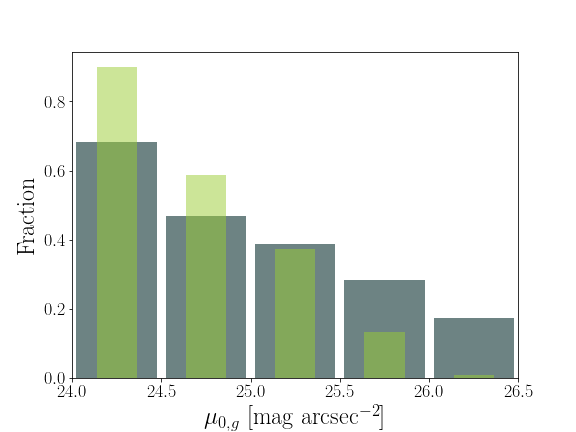}
    \caption{\textit{Distribution of matched sources with \cite{Tanoglidis+2021} vs. central surface brightness. The darker histogram represents the distribution of surface brightnesses in the SMUDGes catalog. The lighter histogram, with thinner bars, represents the distribution of surface brightnesses among matched sources. The Figure shows that the matched source distribution is skewed toward higher central surface brightness, demonstrating that the fainter objects in SMUDGes are underrepresented in the \cite{Tanoglidis+2021} catalog.}}
    \label{fig:tangolidis_comp}
\end{figure}

SMUDGes has the greatest overlap with the \cite{Tanoglidis+2021} catalog, where we now identify 1261 objects in common. However, that catalog has broader selection criteria and includes many objects with brighter central surface brightnesses than SMUDGes, so care is still warranted before comparing results. Furthermore, as described in Paper II and confirmed here, SMUDGes is especially more sensitive for objects with $\mu_{0,g} > 25.5$ mag arcsec$^{-2}$ (Figure \ref{fig:tangolidis_comp}).

\begin{deluxetable*}{lrr}
\caption{The Catalog$^a$}
\label{tab:catalog}
\tablehead{
\colhead{Column Name}&
\colhead{Description}&
\colhead{Format and/or Units}\\
}
\startdata
SMDG Name&Object Name&SMDG designator plus coordinates\\
RAdeg&Right Ascension (J2000.0)&decimal degrees\\
DEdeg&Declination (J2000.0)&decimal degrees\\
Re&non-circularized effective radius &angular (arcsec)\\
E\_Re&effective radius 1$\sigma$ upper uncertainty&angular (arcsec)\\
Re-bias&effective radius measurement bias&angular (arcsec)\\
e\_Re&effective radius  1$\sigma$ lower uncertainty&angular (arcsec)\\
f\_Re&effective radius uncertainty model flag&0 = good, 1 = extrapolated\\
b/a&axis ratio (minor/major)&unitless\\
E\_b/a&axis ratio 1$\sigma$ upper uncertainty&unitless\\
b/a-bias&axis ratio measurement bias&unitless\\
e\_b/a&axis ratio 1$\sigma$ lower uncertainty&unitless\\
f\_b/a&axis ratio uncertainty model flag&0 = good, 1 = extrapolated\\
n&S\'ersic index&unitless\\
E\_n&S\'ersic index 1$\sigma$ upper uncertainty&unitless\\
n-bias&S\'ersic index measurement bias&unitless\\
e\_n&S\'ersic index 1$\sigma$ lower uncertainty&unitless\\
f\_n&S\'ersic index uncertainty model flag&0 = good, 1 = extrapolated\\
PA&major axis position angle&defined to be [$-$90,90) measured \\
&&N to E, in degrees\\
E\_PA&major axis position angle 1$\sigma$ upper uncertainty&degrees\\
PA-bias&major axis position angle measurement bias&degrees\\
e\_PA&major axis position angle 1$\sigma$ lower uncertainty&degrees\\
f\_PA&major axis position angle uncertainty model flag&0 = good, 1 = extrapolated\\
mu0$X$&central surface brightness in band $X$ ($X \equiv$ g,r,z)&AB mag arcsec$^{-2}$\\
E\_mu0$X$&central surface brightness 1$\sigma$ upper uncertainty in band $X$&AB mag arcsec$^{-2}$\\
mu0$X$-bias& central surface brightness measurement bias in band $X$&AB mag arcsec$^{-2}$\\
e\_mu0$X$&central surface brightness 1$\sigma$ lower uncertainty in band $X$&AB mag arcsec$^{-2}$\\
f\_mu0$X$&central surface brightness uncertainty model flag in band $X$&0 = good, 1 = extrapolated\\
$X$mag&total apparent magnitude in band $X$&AB mag\\
E\_$X$mag&total apparent magnitude 1$\sigma$ upper uncertainty in band $X$&AB mag\\
$X$mag-bias&total apparent magnitude measurement bias in band $X$&AB mag\\
e\_$X$mag&total apparent magnitude 1$\sigma$ lower uncertainty in band $X$&AB mag\\
f\_$X$mag&total apparent magnitude uncertainty model flag in band $X$&0 = good, 1 = extrapolated\\
VC&visual classification flag&0 = good, 1 = rejected,\\
&&2 = observers disagreed\\
SFD&Optical depth at SMDG location (see text)&unitless\\
A$X$mag&Corresponding extinction at SMDG location in band $X$&AB mag\\
comp&Fractional completeness for similar UDGs&unitless\\
f\_comp&Completeness model flag&0 = good, 1=extrapolated, \\
&&2=biases extrapolated\\
\enddata
\tablenote{The catalog is available as the electronic version of this Table. {The uncertainty range is determined by applying the given upper and lower uncertainties to the measured value (by adding the upper uncertainty value and subtracting the lower uncertainty value) but represent the uncertainty range about the bias corrected value.}}
\end{deluxetable*}

\section{Distance by Association}
\label{sec:distances}

We approach the UDG candidate distance challenge in a similar way as previous investigators have. We identify those UDG candidates projected on overdense physical systems, as defined using galaxies with existing redshift measurements, and presume that the UDG candidates are members of that overdensity. This approach was first adopted by \cite{vanDokkum+2015} for candidates projected on the Coma cluster and has been used widely \citep[e.g.,][]{Mihos+2015,Yagi+2016,Trujillo+2017,Roman+2017,vanderBurg+2016,Tanoglidis+2021}. At least for those UDGs projected onto the Coma cluster, this approach has been verified to work for most candidates using spectroscopic follow-up \citep{vanDokkum+2016,Kadowaki+2017,Kadowaki+2021}. The results are likely to become more questionable at lower overdensities and will depend on the degree of correlation between `normal' galaxies and UDG candidates.

Rather than applying this method only in highly overdense systems, e.g., the Coma or Virgo clusters, we aim to expand the environments in which we can provide estimated redshifts. We define the overdensities ourselves rather than depending on existing group or cluster catalogs. 
Our sample of `normal' galaxies with measured redshifts comes from the compilation provided by the on-line database SIMBAD \citep{simbad}.
For each galaxy of interest without a spectroscopic redshift, we draw galaxies from SIMBAD that are projected within 5$^\circ$ and have $0.0017 < z < 0.05$. The lower redshift limit, corresponding to $cz \sim 500$ km sec$^{-1}$, is set to avoid confusion with Galactic sources and, in practice, also removes a few local galaxies from consideration. Although this selection may prevent us from identifying a local UDG, working in a regime where peculiar velocities are comparable to the expansion velocity renders any association and inferred distance highly suspect. Furthermore, our completeness for local UDGs is extremely low (\S\ref{completeness}). The upper limit, $z=0.05$, is guided by our understanding that UDGs larger than $r_e > 6$ kpc are quite rare \citep{vanDokkum+2015,Yagi+2016} and often spurious \citep{Kadowaki+2021}. 

To have roughly equal sensitivity to structures across the relevant volume, we limit ourselves to using bright galaxies, $M_B$ or $M_g < -19$ mag. We include in our consideration only such galaxies in SIMBAD that lie at a projected physical distance less than $r_{C,proj}$ from the galaxy of interest, using the angular diameter distance at the redshift of the SIMBAD galaxy to evaluate $r_{C,proj}$. We explored a range of values for $r_{C,proj}$ ranging from 0.5 to 2.5 Mpc and concluded that 1.5 Mpc provided a compromise between providing an estimated redshift for as many candidates as possible and including contaminating, unrelated overdensities in the line-of-sight galaxy distribution. In practice, the results were not highly sensitive to the choice of $r_{C,proj}$ within this range.

\begin{figure*}
    \centering
    \includegraphics[scale=0.65]{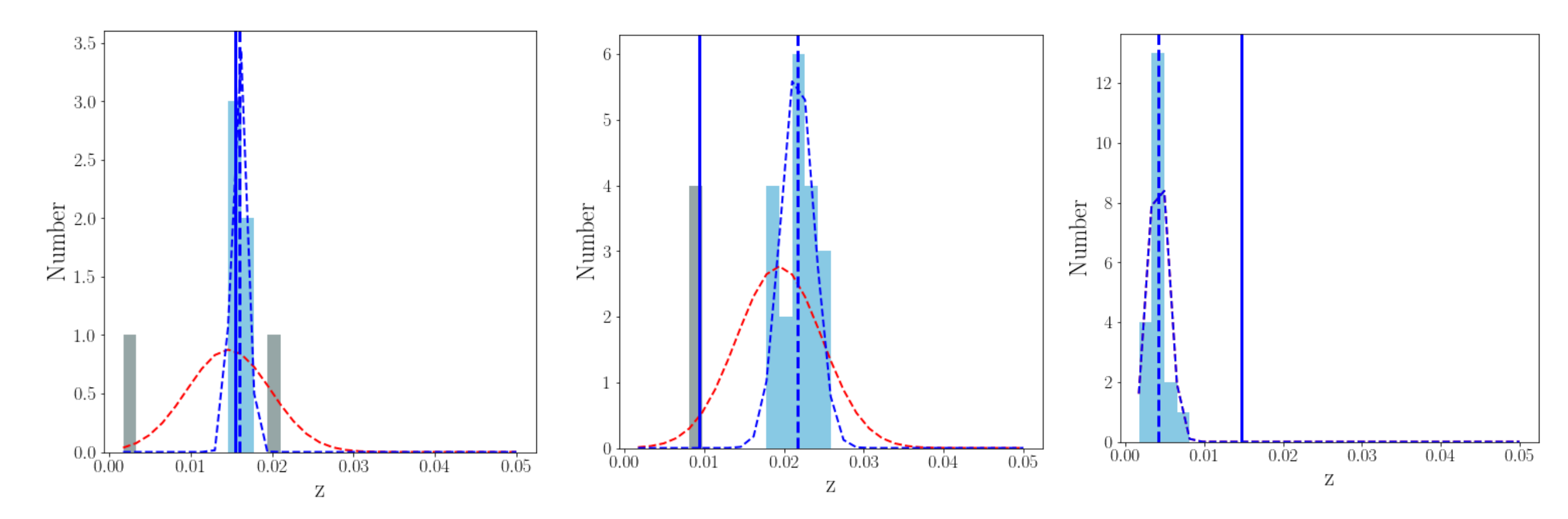}
    \caption{\textit{Example of failure cases for three candidate UDGs. Each panel shows the line-of-sight galaxy distribution, as selected for using the description in the text. The dashed blue vertical line shows the calculated mean redshift of the main concentration of galaxies identified along the line of sight and the solid vertical line the spectroscopic redshift of the UDG candidate. The blue histograms highlight the galaxies in the main peak, as defined in the text. The blue dashed curve shows the Gaussian fit to these galaxies, while the red dashed curve shows the fit to all of the galaxies along each line of sight. The left panel shows an example where the estimated and spectroscopic redshifts agree well, but the Gaussian fit to all of the galaxies is much broader, suggesting the possibility for confusion due to other galaxies along the line of sight. Unfortunately, we would reject this estimated redshift even though it is indeed correct. The middle panel shows another case that is likely to be rejected because of the different widths of the fitted Gaussians, but here we see that the estimated redshift is an inaccurate estimate of the spectroscopic redshift. Finally, in the right panel we show a case where the fit is accepted because the full distribution and main peak are identical, but the UDG lies at a redshift with no cataloged galaxies. This is an example of a catastrophic redshift estimation failure that is impervious to our choice of criteria.}}
    \label{fig:examples}
\end{figure*}

Using galaxies that satisfy all of these criteria  surrounding each galaxy of interest, we then drop from consideration those that have 2 or fewer potential companion galaxies. We opted for a minimum of three associated galaxies because we want to measure the mean recessional velocity and  its dispersion for the putative group. 
To help determine whether an association between the UDG candidate and a structure along the line of sight can be plausibly made  we 
divide the galaxies along the line of sight into 30 redshift bins between $0.0017 < z < 0.05$ and identify the location of the peak in the number distribution. 
We fit a Gaussian to the redshift distribution of all the potential companions. We refer to the standard deviation of this distribution as $\sigma_{ALL}$. Then
we work our way to both lower and higher redshift bins from the peak of the distribution until there are zero galaxies in a bin. We fit a second Gaussian to this trimmed $z$ distribution and refer to it as $\sigma_{TRIM}$. The ratio between these two measurements is the measure of the complexity of the line of sight galaxy distribution that we will use to accept or reject an estimated redshift.

We explore this and other potential criteria we might use to help us discriminate between correct and incorrect estimated redshifts using two different sets of galaxies. First, we estimate redshifts for  galaxies with spectroscopic redshifts drawn at random from SIMBAD to hone the method and evaluate the resulting estimated redshifts. The advantage of this sample is that it can be quite large, limited only by the number of local galaxies in the SIMBAD database. The disadvantage is that these galaxies do not match the properties of UDG candidates and so may provide somewhat skewed results. Second, 
we estimate redshifts for a sample of SMUDGes sources with previously-obtained spectroscopic redshifts. We compile this sample by searching for cataloged sources projected within 6\arcsec\ of our candidates and with $-300 < cz/{\rm km\  sec}^{-1} < 30000$ in NED, SIMBAD, and SDSS. The association is fairly secure because the projection of a low redshift source that is sufficiently bright to have been previously targeted for spectroscopy would have most likely prevented us from detecting the UDG candidate and we have visually examined each SMUDGes candidate during our visual classification process. We complement this list with the 68 redshifts from our own set of compiled spectroscopic measurements from a variety of sources \citep{Kadowaki+2021}. The principal disadvantage of the spectroscopic UDG sample is its size, with only 187 sources, although it too can be somewhat non-representative given the heterogeneous selection of spectroscopic targets across different studies.

Using both of these samples, we now aim to understand under what conditions the estimated redshifts are most likely to be accurate. In Figure \ref{fig:examples} we show three example of redshift estimation failures using SMUDGes sources with existing spectroscopic redshifts. In the first two cases we show how a difference between $\sigma_{ALL}$ and $\sigma_{TRIM}$ reflects additional structure along the line of sight, including one case (middle panel) where that structure leads to a catastrophic redshift estimate. In the left panel, we see that we unfortunately reject a case where the estimated redshift was correct due to a complex line of sight. In the right panel of the Figure, we show a case where 
$\sigma_{ALL}$ and $\sigma_{TRIM}$ agree, but the estimated redshift is far from the spectroscopic one. Here we have either identified a SMUDGe that is not associated with other galaxies, we are victims of incomplete redshift catalogs, or the spectroscopic redshift is incorrect. The last option may appear unlikely, but with such low surface brightness objects the spectra are often of very low S/N \citep[e.g.,][]{Kadowaki+2017}. Assuming that the spectroscopic redshift is correct, then this example demonstrates why this approach will never yield 100\% accurate redshift estimates, although it will improve significantly with the increased sampling eventually provided by DESI.

Using randomly selected normal galaxies and defining an accurate estimated redshift as one that is within $3\sigma$ of the spectroscopic redshift, we 
experimented with different thresholds of  $\sigma_{ALL}/\sigma_{TRIM}$. 
We adopt $\sigma$ as measured for the associated group unless $\sigma < 100$ km sec$^{-1}$, in which case we set $\sigma = 100$ km sec$^{-1}$ for this purpose.
For a sample of 10000 randomly selected galaxies with spectroscopic redshifts, we were able to find overdensities to associate them with in 4021 cases. However, the estimated redshift was accurate in only 49\% of those. Restricting the sample to those with unambiguous lines of sight, done by 
requiring $\sigma_{ALL}/\sigma_{TRIM} < 1.5$, increases the accuracy to 72\%. This cut results in a sample size of 734, from the original sample of 4021 with estimated redshifts. 

To see if we can realize any further improvement in the redshift accuracy, we then applied the Scikit-learn random forest classifier \citep{scikit-learn} with $\sigma_{ALL}$, $\sigma_{TRIM}$, the number of galaxies in the associated peak, the fraction of galaxies along the line of sight in that associated peak, and the estimated redshift as the relevant features. We find an improvement in the accuracy to 92\%, while now having estimated redshifts for only 556 galaxies. The combined procedure therefore yields estimated redshifts for $\sim$ 6\% of this sample, although with reasonably high accuracy (estimated to be $\sim$ 90\%). Bypassing the initial cut on $\sigma_{ALL}/\sigma_{TRIM}$ and relying only on the random forest classifier results in lower accuracy.

\begin{figure*}
    \centering
    \includegraphics[scale=0.31]{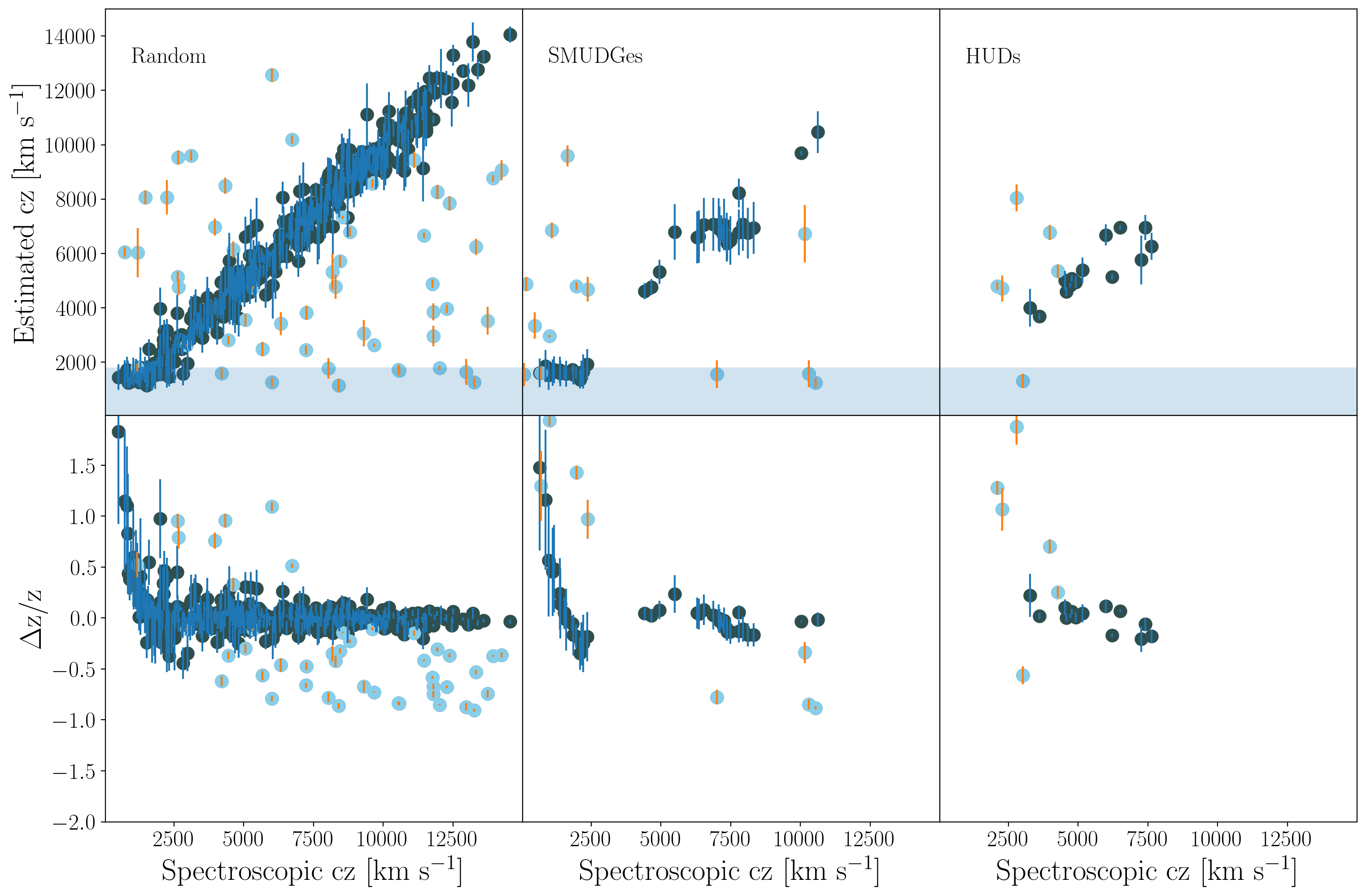}
    \caption{\textit{Comparison of spectroscopic and estimated redshifts. Darker symbols represent those objects for which the estimated redshift is within 10\% or 3$\sigma$ of the spectroscopic redshift, while the lighter symbols for those where it is not. In the left panel we show the results for galaxies drawn randomly from the SIMBAD database, in middle panel for UDGs with spectroscopic redshifts, and in the right panel for a sample of HUDs with spectroscopic redshfits. Shaded regions show exclusion zones based on estimated redshifts that eliminate where $cz$ is too low to yield a reliable distance.}}
    \label{fig:z_comparison}
\end{figure*}

We now proceed to determine if this accuracy is reproducible with the actual UDG candidates.
Of the 187 candidates with spectroscopic redshifts, we are able to associate overdensities along the line of sight for 130. Of these, only 57 are considered as reliable based on the $\sigma_{ALL}/\sigma_{TRIM} < 1.5$ criterion. Applying the further random forest classifier results in a final sample of 52 candidates with estimated redshifts and the resulting accuracy is 76\%. We find a far larger return rate of redshift estimates, but a lower accuracy percentage in comparison to the random galaxy sample. 
We attribute both of these aspects to a feature of this sample that can be seen in the middle panel of Figure \ref{fig:z_comparison}. A large fraction of the UDG candidate sample with spectroscopic redshifts lie either in the Virgo or Coma cluster regions. This is simply the result of spectroscopic campaigns preferentially targeting those areas \citep[cf.][]{Kadowaki+2017,chilingarian}, but it does lead to both a higher redshift return rate, because candidates lie in a region of sky with overdensities, and a lower accuracy because any candidate projected onto the Virgo or Coma clusters will have the corresponding redshift estimate whether or not it lies in either cluster. 

Finally, we also test our method using a distinct sample of low surface brightness galaxies, an H{\small I}-bearing sample of ultra-diffuse galaxies 
\citep[HUDs;][]{Leisman+2017,Janowiecki+2019}. These are likely to be the most isolated subset of UDGs and, as such, pose the greatest challenge to our redshift estimation technique. Because of this relative isolation, the redshift yield (8\%) is lower than that for the SMUDGes spectroscopic set (28\%), but it is nevertheless similar to that of the random sample (7\%), while the accuracy (74\%; see \S\ref{sec:understanding}) is comparable to that of the SMUDGes spectroscopic set (76\%; see \S\ref{sec:understanding}). We conclude that across the range of prospective UDG properties and environments, our redshift accuracy is likely to be $\sim$ 75\% when the method yields a redshift estimate.

Applying the method and criteria to the full sample of available UDG candidates from SMUDGes (5598) we are able to estimate redshifts for 1079. The return rate is 19\% and, as expected, lies between the return rates for the random galaxy and spectroscopically-confirmed UDG samples. In the end, the random forest classifier removes about 10\% of the estimated redshifts in the SMUDGes training sample, the HUD sample, and the full SMUDGes catalog. Thus, it provides at best a modest improvement given the estimated 25\% contamination rate, but it does highlight a potential way forward once the training samples are larger.

\begin{figure}
    \centering
    \includegraphics[scale=0.4]{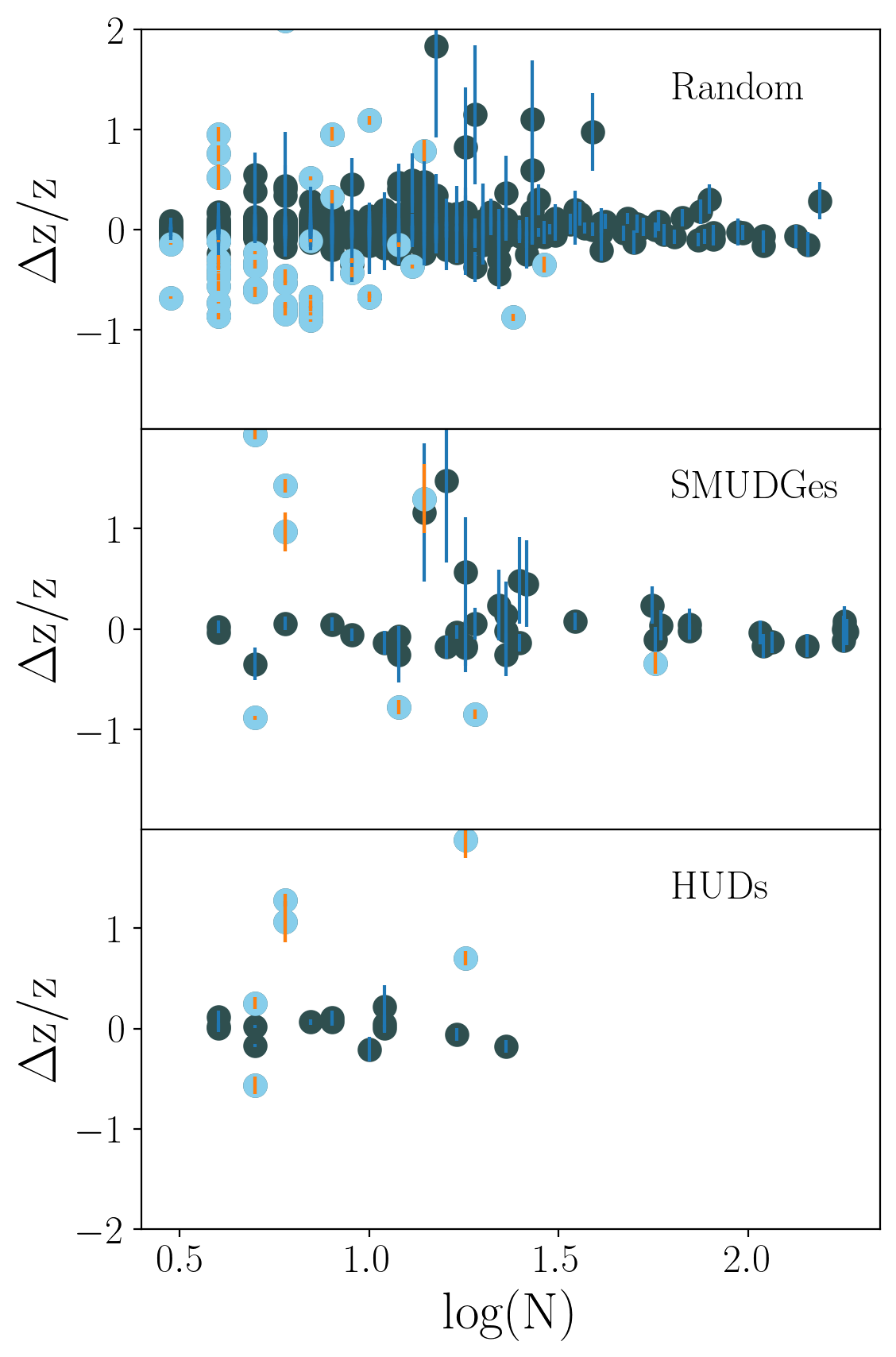}
    \caption{\textit{Redshift errors vs. the number of galaxies in the associated overdensity. Symbols coded as in Figure \ref{fig:z_comparison}.}}
    \label{fig:deltaz_n}
\end{figure}

\subsection{Understanding the Redshift Estimates}
\label{sec:understanding}

Figure \ref{fig:z_comparison} is our starting point for the discussion of the properties of the estimated redshifts. As illustrated by the left and middle panels, the method appears to produce large fractional errors for galaxies at low $cz$. For the SMUDGe data this is mostly due to Virgo members, which have a large intrinsic velocity scatter and are at relatively low $cz$, leading to large fractional errors. This is related to the generic problem of using recessional velocities, which include peculiar velocities, to estimate distances for nearby objects. We exclude objects with an estimated $cz < 1800$ km sec$^{-1}$ in our discussion of physical properties (\S\ref{distance_dependent}). That cut is shown in the shaded region of the upper set of panels in Figure \ref{fig:z_comparison} and is constructed primarily to exclude very nearby groups and clusters. Other than this one set of outliers, the method appears to be roughly equally capable across the relevant redshift range. The method, after this cut, yields accurate redshifts for 76\% and 74\% of the SMUDGes and HUD samples, respectively. Note that even the majority of the Virgo candidates are likely to have been assigned the correct distance, as they are probably truly in Virgo, but given the large angular extent of Virgo there are also a significant number of accidental projections and we opt to exclude the entirety of these low $cz$ objects from our final sample.

In Figure \ref{fig:deltaz_n} we examine the estimated redshift quality vs. environment. As might be expected and appears evident in the randomized galaxy sample (upper panel), the estimated redshift accuracy and precision are both lower for overdensities with a small number of associated galaxies, $N$. While the same may be true for the SMUDGes and HUDs samples, it is not clearly the case, perhaps due to small number statistics. Because it is of scientific interest to explore the distribution of UDGs in low $N$ environments, we choose to not require larger $N$  values than we currently do (i.e., $N \ge 3$). However, we do recommend additional caution when using the estimated redshifts for systems associated with $N \le 6$ environments. Of our full sample of estimated redshifts, only about 20\% are associated with environments of $N \le 6$. 

\begin{figure}
    \centering
    \includegraphics[scale=0.4]{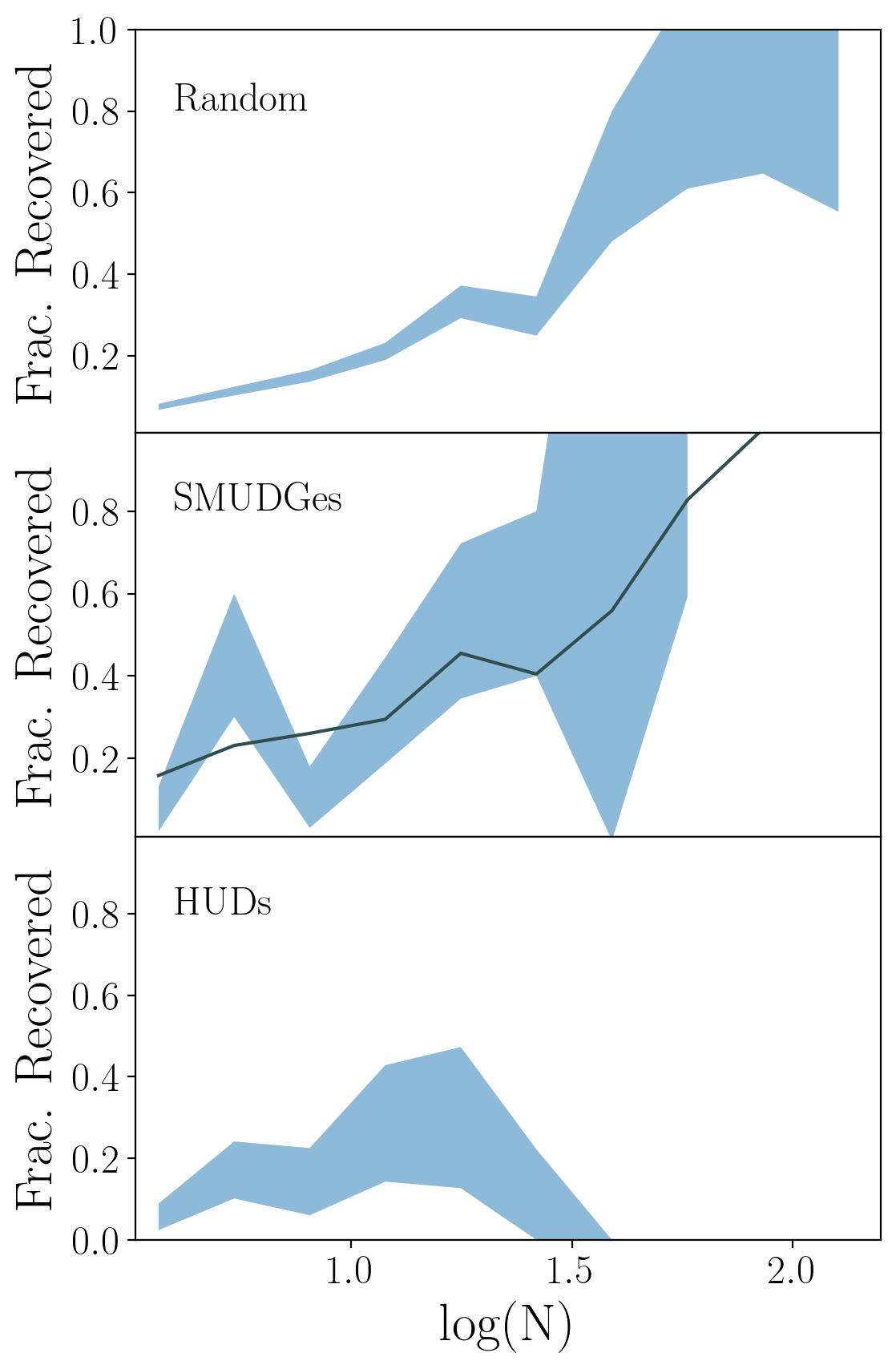}
    \caption{\textit{The fraction of galaxies with recovered estimated redshifts as a function of the environment they are associated with by our technique. As expected, we tend to accept the redshift association more often in richer systems. Also as anticipated, the HUD sample avoids rich systems. The recovered redshifts for the full SMUDGes sample (solid line in middle panel) follows the trend established with the spectroscopic sample, showing no clear bias in the environments that are sampled relative to the training sets. Shaded regions represent 1$\sigma$ confidence bounds estimated using Poisson errors.}}
    \label{fig:recovered_fraction_N}
\end{figure}

\begin{figure}
    \centering
    \includegraphics[scale=0.4]{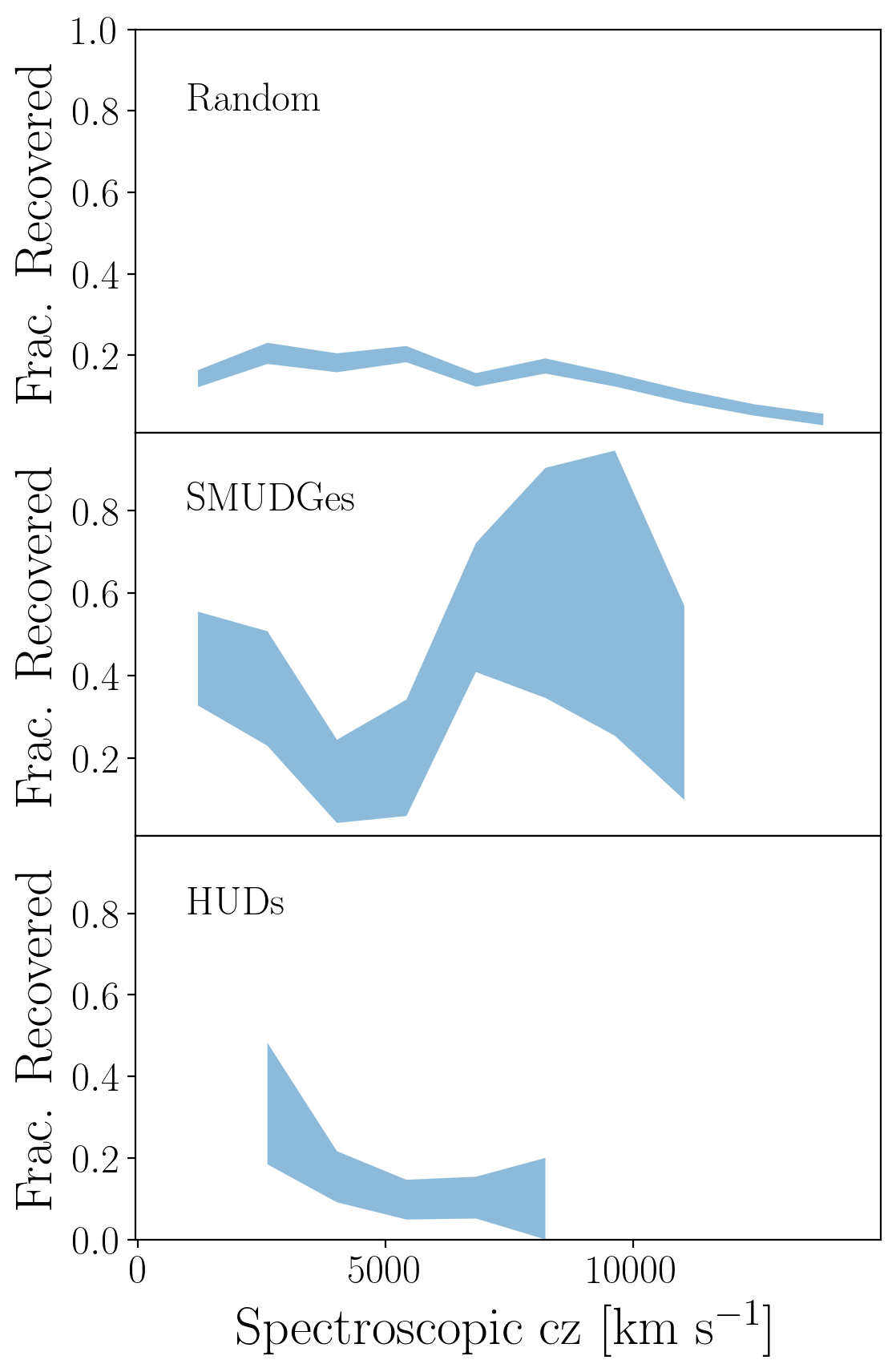}
    \caption{\textit{The fraction of galaxies with recovered estimated redshifts as a function of the spectroscopic redshift. We find little dependence of the recovery fraction on redshift across the relevant redshift range for the random and HUD samples. The SMUDGes sample shows strong fluctuations but these map the influence of the Virgo and Coma clusters on this particular data set. Shaded regions represent 1$\sigma$ confidence bounds estimated using Poisson errors.}}
    \label{fig:recovered_fraction_spec}
\end{figure}

The potential increased uncertainty for systems associated with low $N$ suggests that we examine how the systems with recovered $cz$ are distributed in $N$. We see in Figure \ref{fig:recovered_fraction_N} that we are much more successful at recovering systems in high density environments. This is not unexpected because one would expect to assign systems projected near a rich system to that system with relatively little confusion, e.g., the Coma cluster, and have a more difficult time assigning candidates to those projected near poor systems. As such, the sample of UDGs with estimated redshifts is not representative of the overall sample in terms of the numbers of UDGs per environment, but there is no evidence (yet) that it is not representative in terms of other UDG properties. The recovery fraction as a function of $N$ is consistent between our SMUDGes spectroscopic sample and the full sample (solid line in middle panel of Figure \ref{fig:recovered_fraction_N}). We find little variation in the recovery fraction with spectroscopic redshift, except for the features introduced by the Virgo and Coma clusters in the SMUDGes spectroscopic sample (Figure \ref{fig:recovered_fraction_spec}). 

\begin{figure}
    \centering
    \includegraphics[scale=0.3]{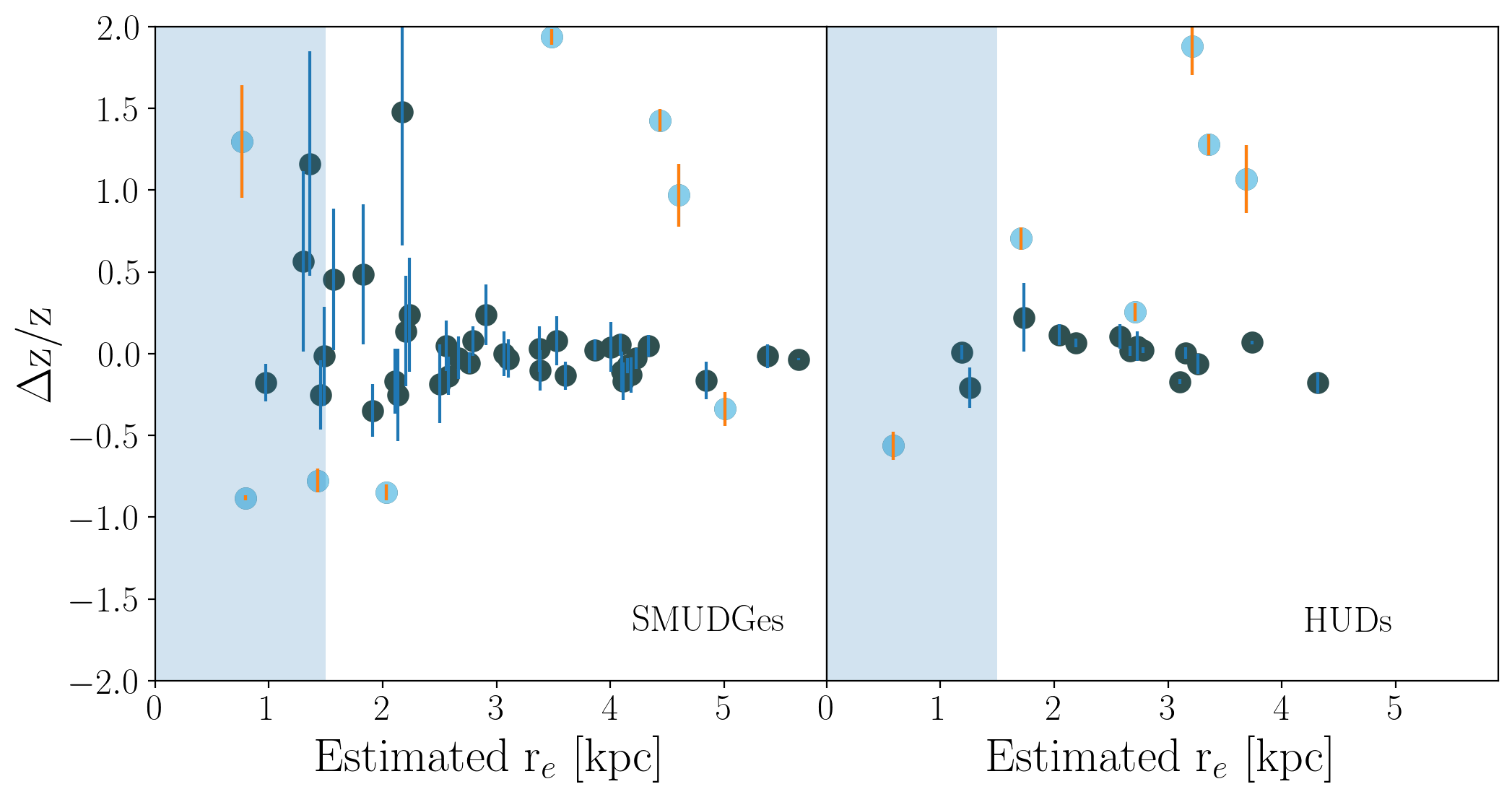}
    \caption{\textit{The estimated redshift errors as a function of inferred $r_e$ in physical units. Points coded as in Figure \ref{fig:z_comparison}. Shaded region shows exclusion region for UDGs. Catastrophic errors are scattered along $r_e$ and scatter among successful redshift estimates is independent of $r_e$.}}
    \label{fig:check_re}
\end{figure}

Finally we examine the estimated redshift quality as a function of a derived parameter, physical $r_e$. We have used the physical value of $r_e$ partly as a prior in our methodology by not allowing redshift solutions that result in what we considered to be unphysically large values of $r_e$ (i.e., $r_e \ge 6$ kpc). We examine the resulting distribution for the two UDG samples in Figure \ref{fig:check_re}. We find that the recovered redshift behaves well over the range of inferred sizes that are consistent with a UDG classification (i.e., $r_e > 1.5$ kpc). We find that no further selection on $r_e$ would improve the estimated redshifts. 

We conclude that while the estimated redshifts do not provide redshifts for a representative subset of the UDG candidates in terms of the numbers across environment, in all other respects they behave as needed to provide a set of redshifts that are likely to be correct in $\sim$ 75\% of the cases and have no 
apparent biases in terms of redshift or environment. For cases that match the criteria that we will use to select the sample of ``redshift-confirmed" UDGs ($\sigma_{ALL}/\sigma_{TRIM} < 1.5$, $r_e > 1.5$ kpc, $cz > 1800$ km sec$^{-1}$),
the fractional statistical precision on $cz$ for the $\sim$ 75\% of the sample with accurate redshifts is 0.09, 0.12, and 0.11 as calculated from the random, SMUDGes, and HUD samples, respectively. Of course, for the 25\% of the sample with catastrophic redshift estimates the errors turn out to be much larger. Because of this hybrid error distribution it is impossible to propagate errors through a frequentist analysis, although one could forward model these uncertainties for a given model and compare to the data.

\begin{figure}
    \centering
    \includegraphics[scale=0.23]{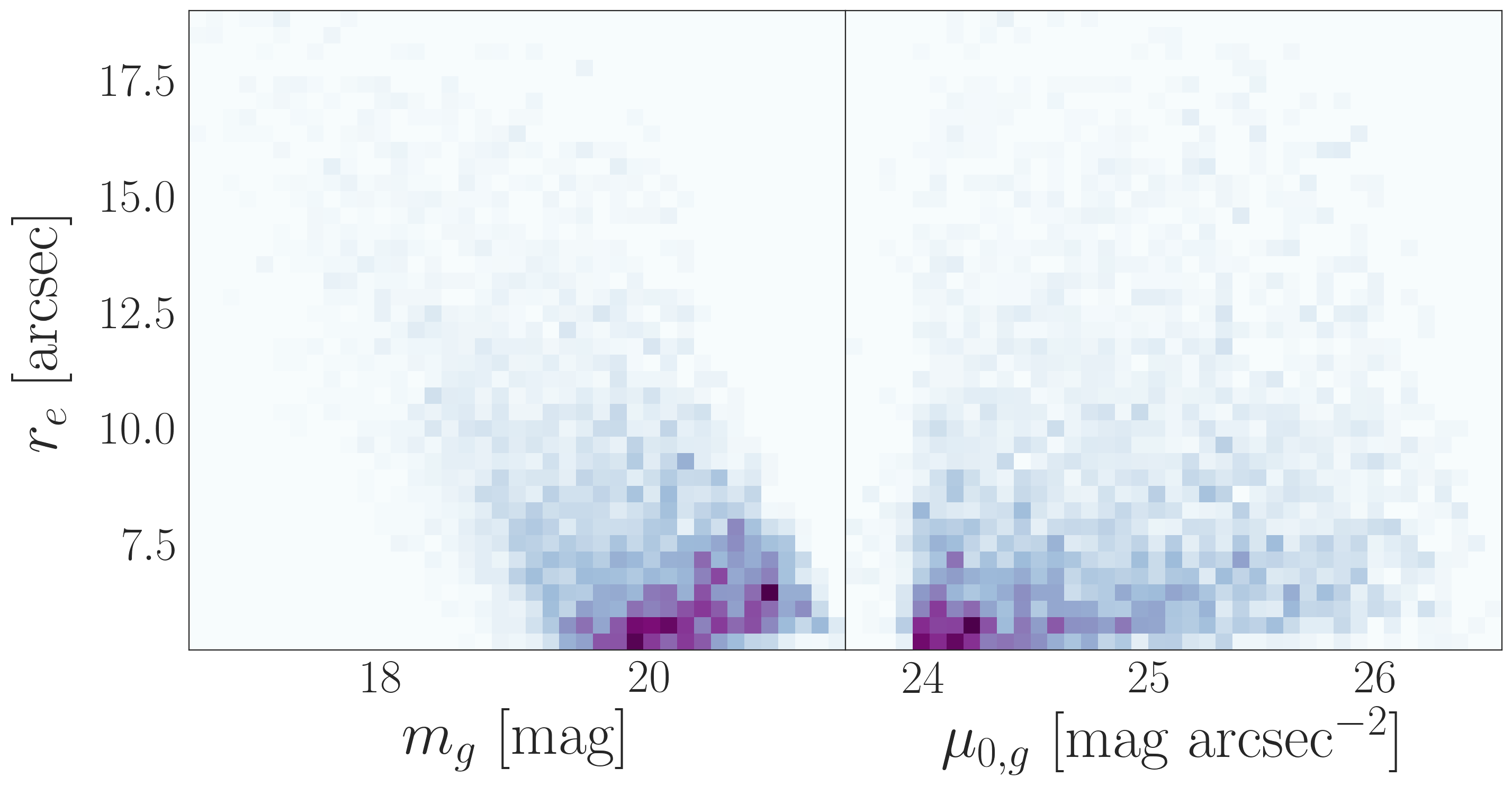}
    \caption{\textit{Distribution of basic observational parameters for the cataloged UDG candidate sample. {A small number of candidates have $\mu_{0,g} < 24$ mag arcsec$^{-2}$ once the measurement is bias corrected.}}}
    \label{fig:intro}
\end{figure}

\section{Results}
\label{sec:results}
In Figure \ref{fig:intro} we present a basic overview of the parameter distribution of the UDG candidates in the catalog by showing the distributions of $r_e$, $\mu_{0,g}$, and $m_g$. The concentrations of candidates toward smaller, higher surface brightness objects are evident. The surface brightness limits which are vertical in the right panel of the Figure, imposed by definition at the bright end and by the nature of the data at the faint end, lead to effective diagonal bounds in the left panel. 

We now describe some preliminary results from an inspection of both the full catalog and the subsample with estimated redshifts. In both cases, we do not consider objects that were visually rejected by at least one reviewer and those which lie in regions of parameter space that are less than 25\% complete. The latter criterion is applied to avoid being misled by a few objects with large, and uncertain, correction factors. 

We will discuss cases where loosening this criterion makes a qualitative difference to the interpretation. These are all cursory demonstration cases for the catalog and more complete analysis and discussion will follow elsewhere.

\begin{figure}
    \centering
    \includegraphics[scale=0.45]{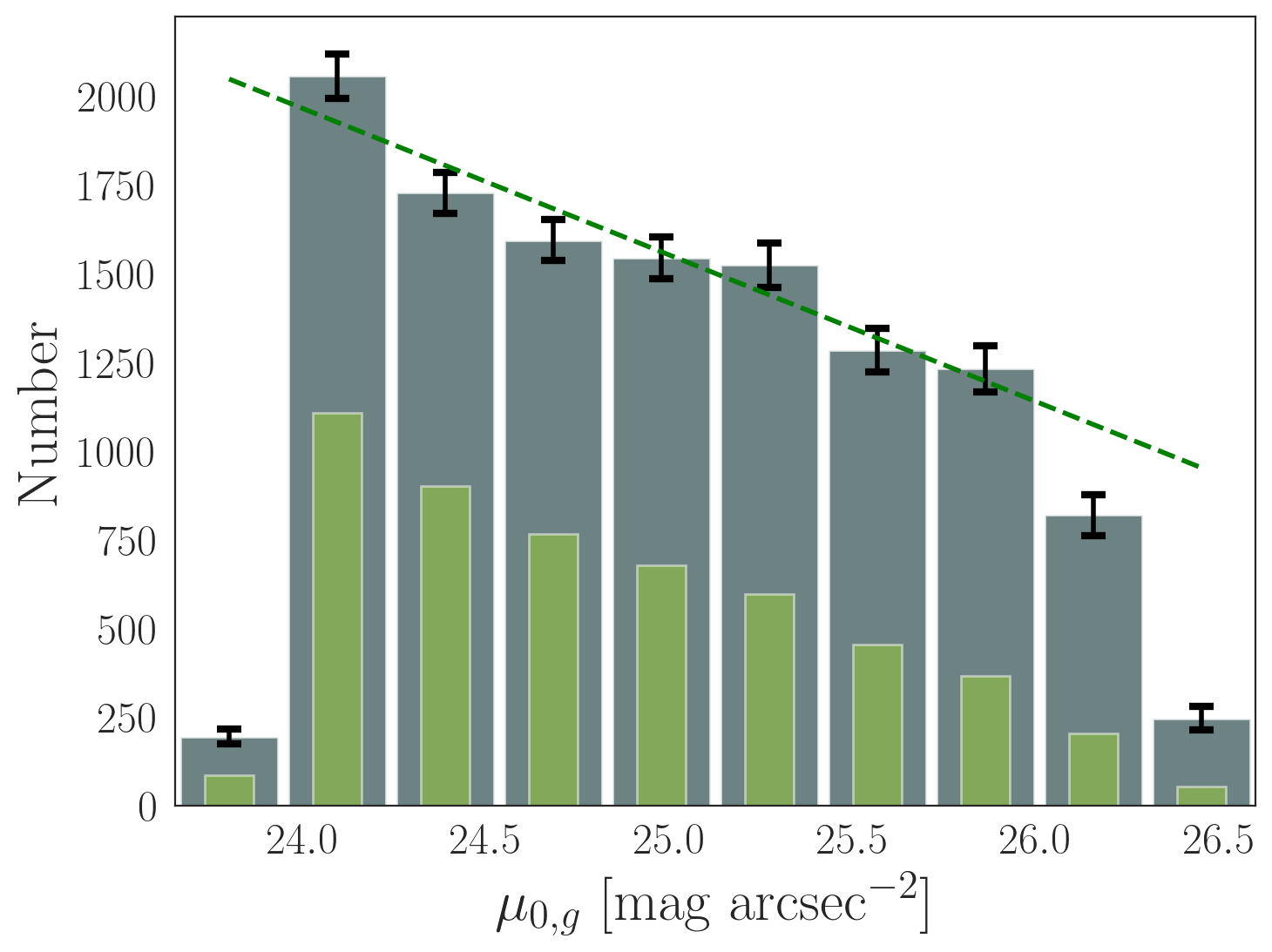}
    \caption{\textit{Surface brightness distribution of sources in the current SMUDGes catalog. The thin green bars represent the raw distribution, while the broader, darker bars represent the completeness corrected distribution. Errorbars represent Poisson noise, multiplied by completeness fraction. The line is a least squares fit to all bins with $>$ 1000 counts. A small number of candidates have $\mu_{0,g} < 24$ mag arcsec$^{-2}$ once the measurement is bias corrected.}}
    \label{fig:surf_brightness_distribution}
\end{figure}

\begin{figure}
    \includegraphics[scale=0.23]{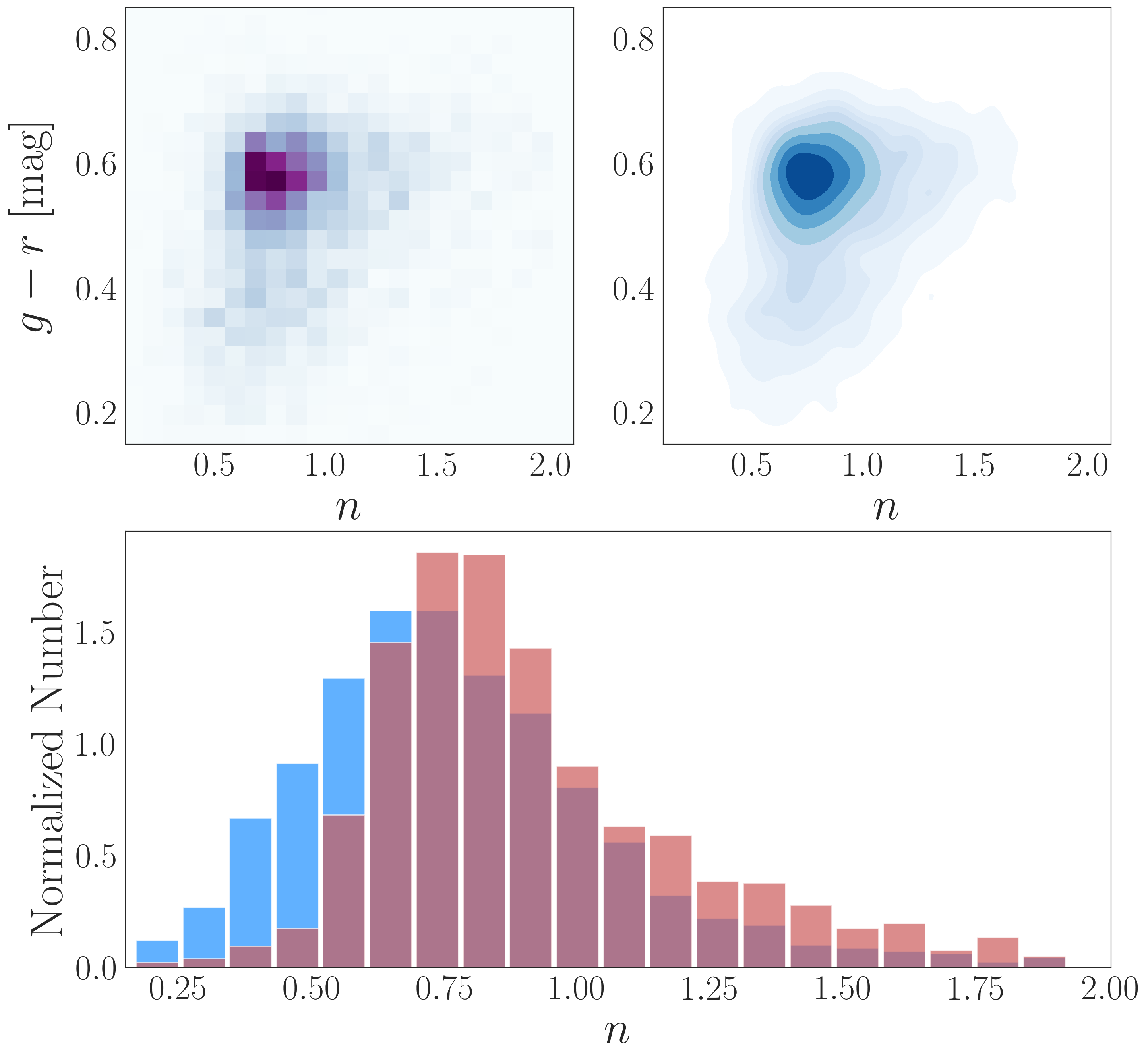}
    \caption{\textit{The  distribution of color vs. S\'ersic $n$ index. The upper left panel shows the completeness-corrected binned distribution while the upper right panel shows the smoothed distribution of the same data. 
    The lower panel shows the distribution of blue candidates ($g-r < 0.45$) in the blue histogram and red candidates ($g-r > 0.55$) in the red histogram. Bluer candidates tend slightly toward smaller values of $n$, while those with the largest $n$ values are red. This large $n$-red wing of the distribution is sparsely populated and highly incomplete, therefore less certain than the blue-small $n$ wing. The sample is dominated by sources with $n \sim 0.8$ and $g-r \sim 0.6$ mag.}}
    \label{fig:n_vs_color}
\end{figure}

\begin{figure}
\centering
    \includegraphics[scale=0.23]{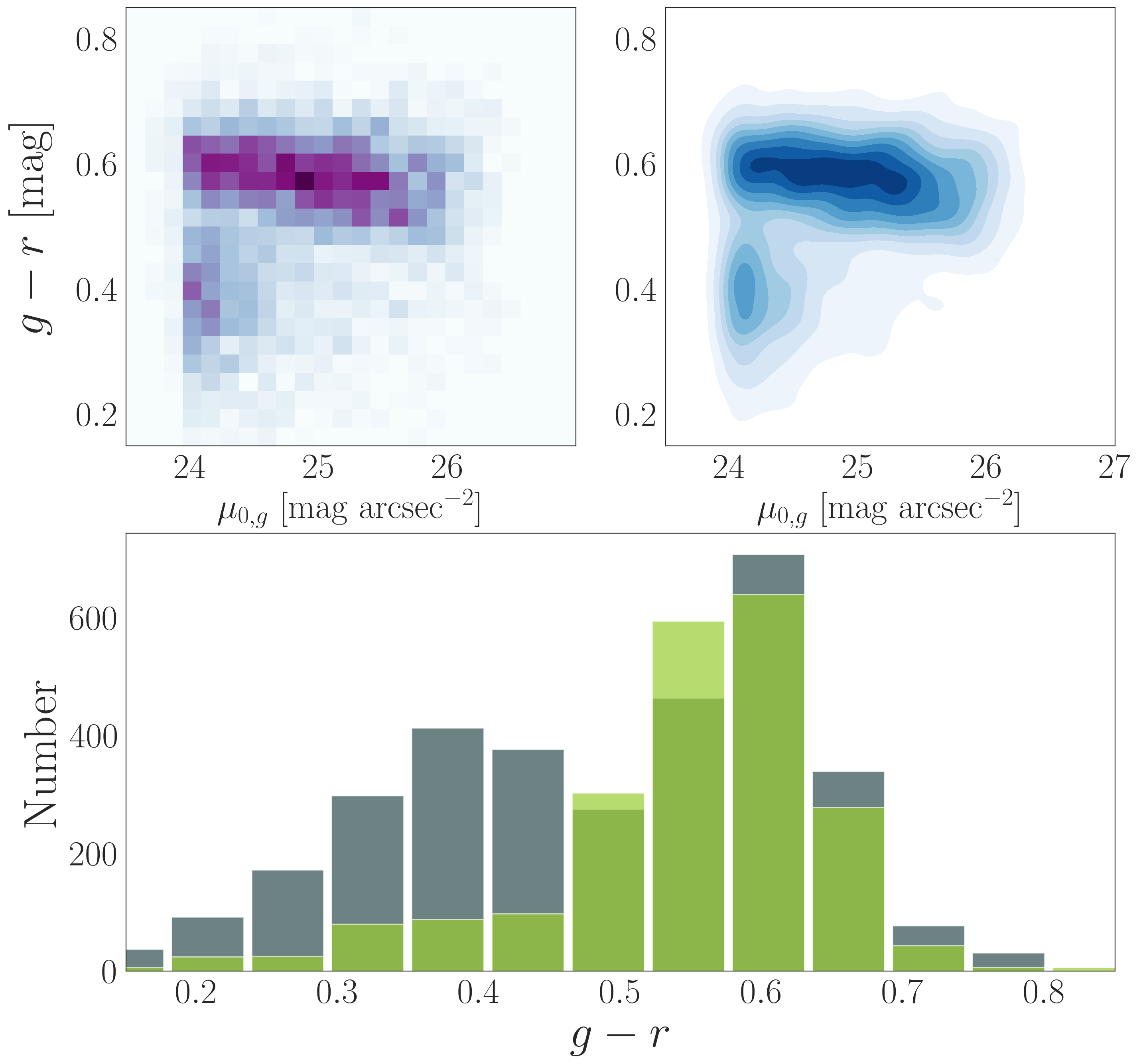}
    \caption{\textit{The distribution of color vs. central surface brightness in the $g$-band. The upper left panel shows the completeness-corrected binned distribution while the upper right panel shows the smoothed distribution of the same data. In the lower panel we compare the distribution of colors for brighter surface brightness ($24 < \mu_{0,g}/{\rm mag\ arcsec}^{-2} < 24.5$) in gray and the somewhat lower surface brightness ($25 < \mu_{0,g}/{\rm mag\ arcsec}^{-2} < 25.5$) candidates in green. Blue UDG candidates are mostly confined to the higher central surface brightnesses. The blue cloud nearly disappears by $\mu_{0,g} \sim 25$ mag arcsec$^{-2}$.
    The red sequence of objects has a slight tilt that we will return to when discussing the color-magnitude relation in \S\ref{sec:color-mag}. A small number of candidates have $\mu_{0,g} < 24$ mag arcsec$^{-2}$ once the measurement is bias corrected.}}
    \label{fig:color_vs_surface_brightness}
\end{figure}

\begin{figure}
    \centering
    \includegraphics[scale=0.23]{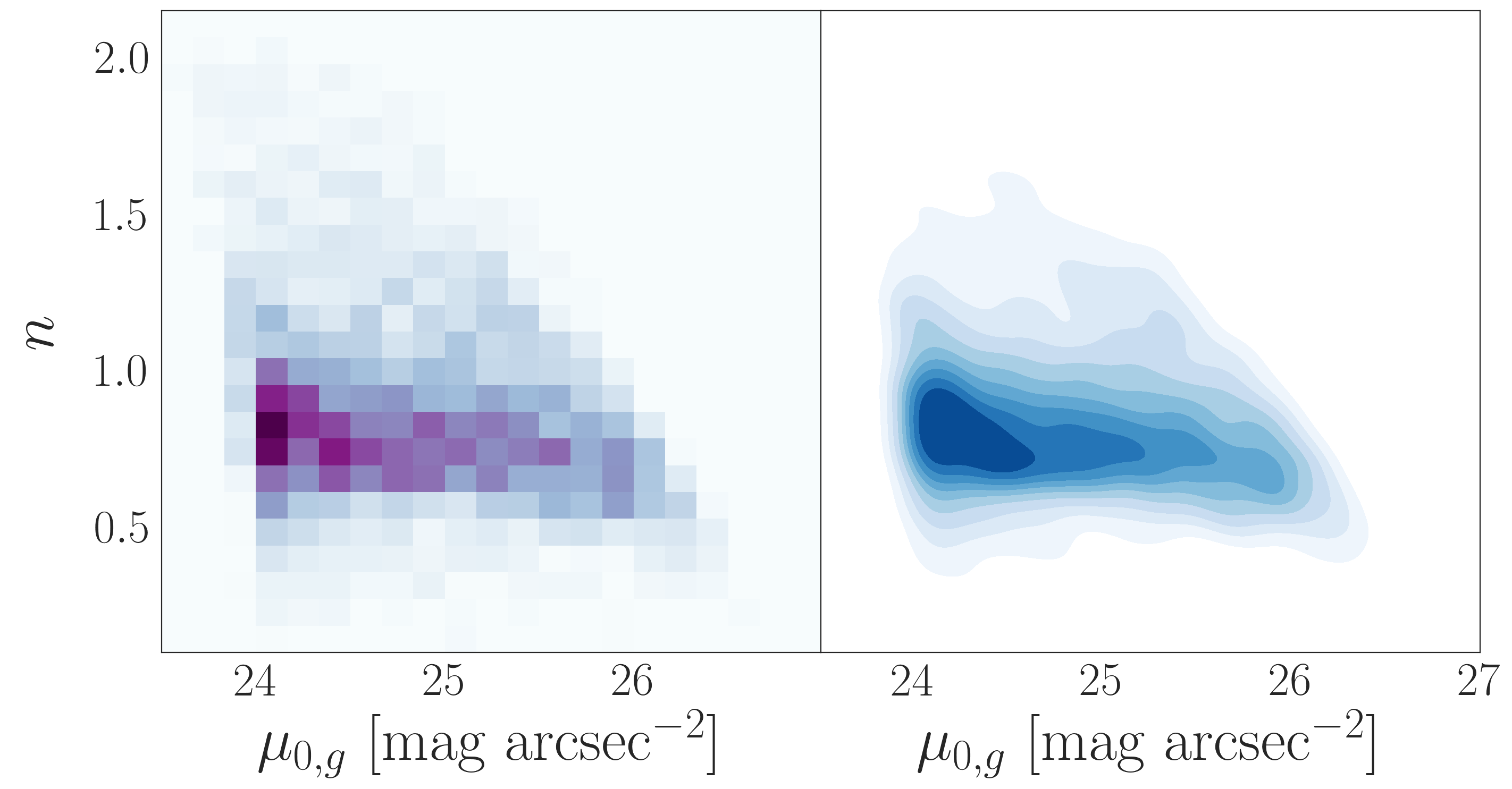}
    \caption{\textit{The 2-D distribution of $n$ vs. central surface brightness in the $g$-band. The left panel shows the completeness-corrected binned distribution while the right panel shows the smoothed distribution of the same data. A small number of candidates have $\mu_{0,g} < 24$ mag arcsec$^{-2}$ once the measurement is bias corrected.}}
    \label{fig:n_vs_surface_brightness}
\end{figure}

\begin{figure*}
    \centering
    \includegraphics[scale=0.3]{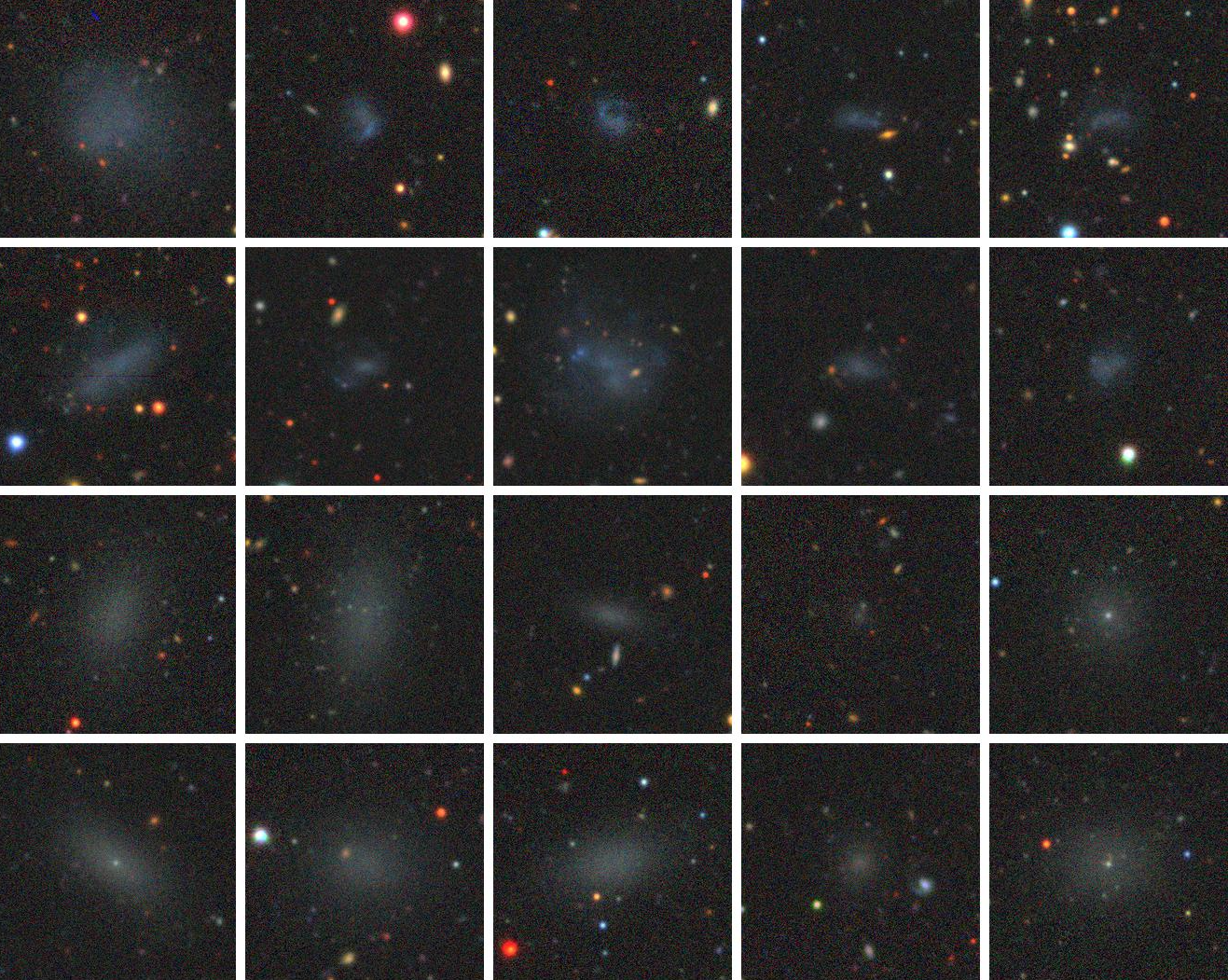}
    \caption{\textit{Examples of UDGs in the blue ($0.2 < g-r < 0.45$) and red ($0.55 < g-r < 0.7$) higher central surface ($\mu_{0,g} < 24.5$) populations. Blue galaxies in the upper two rows, red in the lower two. Some blue UDGs show more internal structure and irregularities. Some red UDGs show what appear to be nuclear star clusters.}}
    \label{fig:red-blues}
\end{figure*}

\subsection{Distance-Independent Results}

With the much larger sample of candidates in hand we now revisit the results we presented in Paper II. First,
we had found no significant decline in the number of UDG candidates as a function of $\mu_{0,g}$ to the limit of our survey ($\sim$ 26.5 mag arcsec$^{-2}$). However, in Figure \ref{fig:surf_brightness_distribution} we now see a significant decline. If we remove from consideration the two faintest bins, as well as the few sources where the measurement bias correction results in $\mu_{0,g} < 24$, a linear fit to the distribution, as shown in the Figure, results in a slope that is 7$\sigma$ away from zero. One concern is that the uncertainties are simply Poisson and do not account for uncertainties in the completeness corrections. Statistically, the uncertainties in the corrections are small as we use many simulated sources to recover the corrections. However, there are possible systematic uncertainties if the nature of the sources differs significantly from what is assumed in those simulations. We have no reason to suspect this is the case, particularly at $\mu_{0,g} \sim 26$ mag arcsec$^{-2}$ where we have many candidates and the sources are well modeled with the same S\'ersic models as we use for brighter sources. However, fainter than this we have both fewer sources and the fitting uncertainties become larger, allowing, perhaps for a different kind of low surface brightness object, for example one that is highly elongated that evades our detection algorithm.
On the basis of the observed distribution for $\mu_{0,g} < 26$ mag arcsec$^{-2}$ and our fit to that distribution, we conclude that there is a decline in the number of candidates as a function of central surface brightness over the range of surface brightness we explore. 

Second, we had found that bluer candidates have smaller S\'ersic $n$. In Figure \ref{fig:n_vs_color}
we confirm that those candidates with the smallest values of $n$ do appear bluer than the bulk of the population and that those with the largest values of $n$ do appear redder. A KS test comparing the $n$ values of blue ($g-r < 0.45$) and red ($g-r > 0.55$) candidates finds that there is less than a $10^{-5}$ chance that they are drawn from the same parent distribution. The bulk of the population has $n \sim 0.8$ across all colors, with differences only at the extremes. Because this difference is only at the extremes, it requires a large sample to see it at all. This statistical difference may explain why previous studies \citep{Greco+2018a,Tanoglidis+2021} did not identify this possible behavior or, as we described in \S\ref{sec:catalog_comp}, this difference may simply reflect that fact that these galaxy samples are all somewhat differently selected.

Third we found that the bluest ($g-r< 0.45$ mag) candidates have $\mu_{0,g} \le 25$ mag arcsec$^{-2}$. That result is more striking with the larger sample available now (Figure \ref{fig:color_vs_surface_brightness}) and we might even move the limit a bit brighter to 24.5 mag arcsec$^{-2}$. A similar trend has been noted before \cite[e.g.,][]{Greco+2018a}.  We argued in Paper II that there is no identified low redshift, blue population that could fade to populate the red sequence below $\sim$26.5 mag arcsec$^{-2}$. This argument appears to remain valid and should be explored further to determine at what surface brightness limit we might expect to only find ``primordial" UDGs. Examples of red and blue UDGs are provided in Figure \ref{fig:red-blues}. 

Lastly, we had found that candidates with fainter $\mu_{0,g}$ tended to smaller $n$. We find that trend here too (Figure \ref{fig:n_vs_surface_brightness}), but the slope of the relationship is small and could be the result of selection biases that are not completely captured in our completeness correction procedure. Indeed we find that the bias observed in our recovered simulated sources is in the same sense as and of larger amplitude than that observed, suggesting that the observed one might be due to a slight undercorrection of the bias. 

\begin{figure}
    \centering
    \includegraphics[scale=0.35]{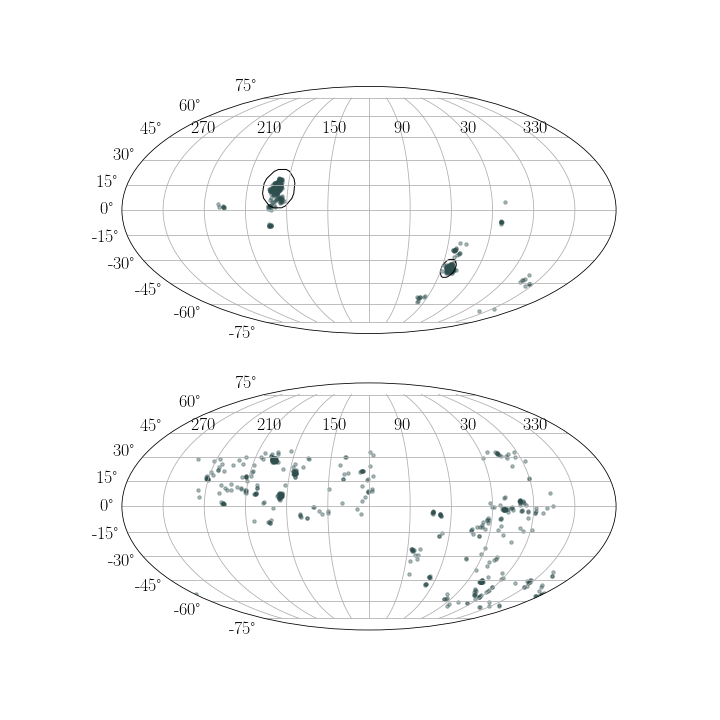}
    \caption{\textit{The distribution of candidates that are estimated to have $r_e < 1.5$ kpc (upper panel) and those that have $r_e > 1.5$ kpc and also have $cz > 1800$ km sec$^{-1}$ (lower panel) so that the estimated distances are more reliable. Ellipses highlight positions of two known nearby ($cz < 1800$ km sec$^{-1}$) clusters (Virgo, upper left; Fornax, lower right). Outside of nearby clusters, we do not find that a large fraction of our candidates are likely to have $r_e < 1.5$ kpc, suggesting that most of our candidates (outside of nearby clusters) are likely to be UDGs (see text for numbers).}}
    \label{fig:skyplot_udg}
\end{figure}

\subsection{Distance-Dependent Results}
\label{distance_dependent}

\subsubsection{The Fraction of UDGs}

A basic question underlying the SMUDGes survey is what fraction of our candidates satisfy the $r_e > 1.5$ kpc criterion.
Of the 1079 with redshift estimates, 679 (63\%) satisfy $r_e > 1.5$ kpc, although 165 of those have $cz < 1800$ km sec$^{-2}$ so we exclude them, leaving us with 514 confirmed UDGs.
However, this fractional return (48\%) is likely to underestimate the return from the full catalog. 
The vast majority of candidates that failed the size criterion lie in two nearby clusters, principally Virgo and Fornax (see upper panel of Figure \ref{fig:skyplot_udg}). Excluding sources with an estimated $cz < 1800$ km sec$^{-1}$, which effectively excludes these two clusters, of the remaining 563 sources 514 (91\%) satisfy the $r_e > 1.5$ kpc criterion. These results are
encouraging and indicate that most SMUDGes candidates are likely to be UDGs. For the following discussion, we present results only for the 514 candidates that satisfy this physical size criterion and $cz > 1800$ km sec$^{-1}$, the latter cut imposed to avoid objects with large distance uncertainties.
We refer to these as our UDG sample.

In Figure \ref{fig:skyplot_udg}, we present the distribution of the UDGs in our sample. Although some of the known clusters are well represented (compare to Figure \ref{fig:skyplot}), many UDGs lie outside the clusters. This distribution suggests that our approach for estimating redshifts is able to extend the technique beyond the strongest overdensities and that the sample of UDGs with redshifts may not be grossly distorted from a general one.

\begin{figure}
    \centering
    \includegraphics[scale=0.23]{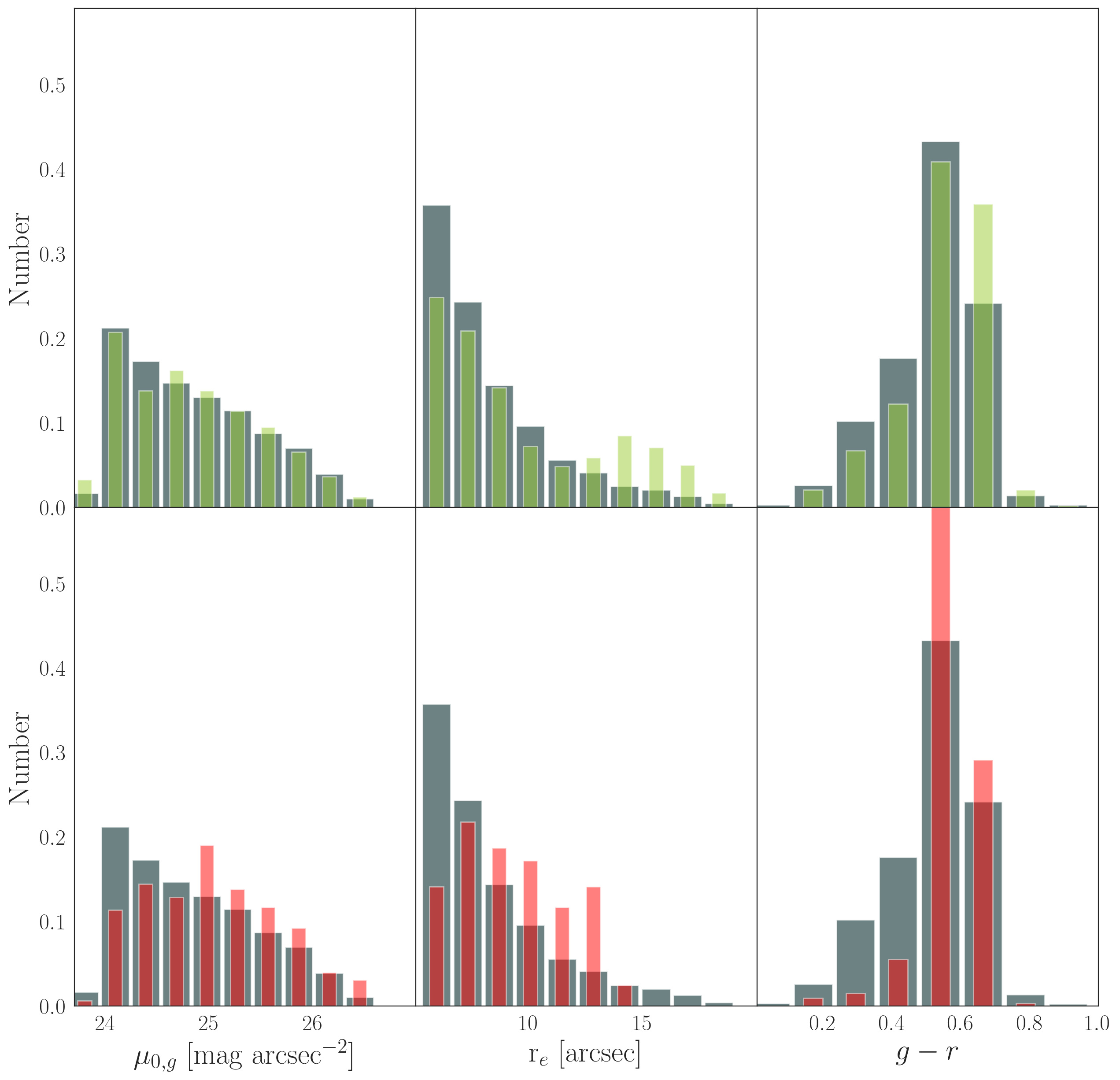}
    \caption{\textit{The comparison of observed properties of the overall sample in grey, the sample of candidates for which we were able to estimate a redshift and the inferred size confirms the candidate as a UDG ($r_e > 1.5$ kpc) in green (upper panels), and candidates with an estimated redshift that are rejected as UDGs ($r_e < 1.5$ kpc) in red (lower panels). All distributions are normalized to sum to unity for comparison.}}
    \label{fig:prop_comparison}
\end{figure}

Next, we support our claim that the bulk of the candidates are more likely to be UDGs than not, by comparing the properties of the candidates as a whole to those of the candidates that are estimated to have $r_e > 1.5$ kpc and those that have $r_e < 1.5$ kpc in Figure \ref{fig:prop_comparison}. The distribution of the UDGs, those with $r_e > 1.5$ kpc, is a closer match to the overall parameter distribution than that of the non-UDGs. The only systematic deviations in the comparison of the UDGs and the overall sample properties are at large angular size, which reflects the over-representation of the nearby clusters in the sample with estimated redshifts, and the slightly redder colors, which is also related to the over-representation of UDGs in denser environments (see \S\ref{sec:understanding}). We conclude that we have no evident reason to suspect that the redshift distribution of those candidates without redshift estimates is grossly different than those for which we obtained estimated redshifts, and therefore that a large number of the remaining candidates will also eventually be spectroscopically confirmed as UDGs.

\subsubsection{Color-Magnitude Relation}
\label{sec:color-mag}

\begin{figure}
    \centering
    \includegraphics[scale=0.23]{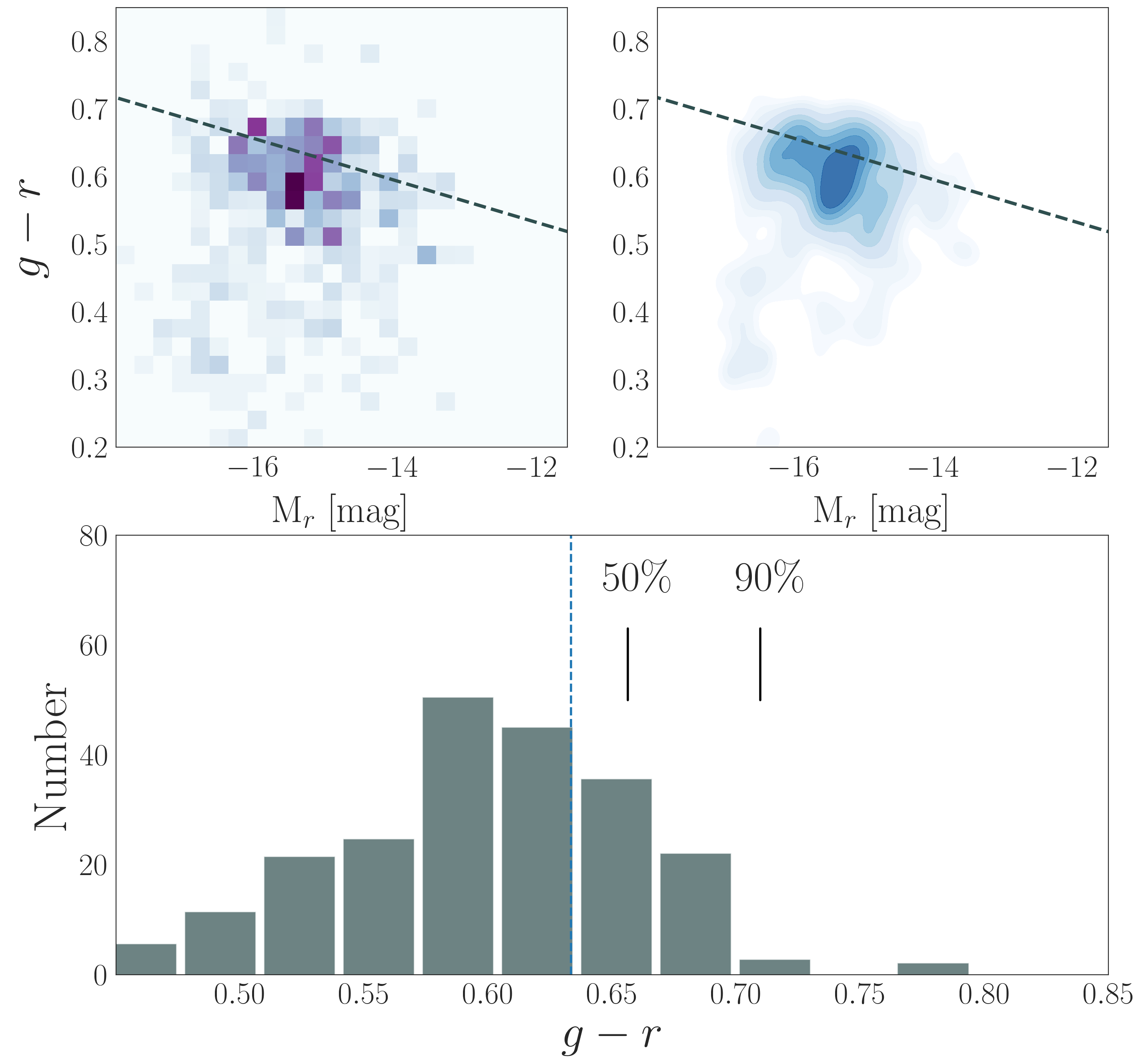}
    \caption{\textit{The color-magnitude diagram for UDGs with estimated redshifts. The distribution has been corrected using our completeness estimates. Upper left panel shows the binned distribution, while the right shows the smoothed distribution. The dashed line is the extrapolation of the red sequence derived for faint ($L < L^*)$ ellipticals \citep{schombert}. In the lower panel we show the color distribution for UDGs with $-15.0 < {\rm M}_r < 15.0$ and note the color along the red sequence of possible progenitors of this population for two different mass loss factors. The dotted vertical line is the position of the red sequence corresponding to M$_r = 15.25$ mag}}
    \label{fig:color_magnitude}
\end{figure}

In Figure \ref{fig:color_vs_surface_brightness} we see a prominent red population of UDG candidates. With the estimated redshifts, we can now place the confirmed UDGs on the color magnitude diagram and determine whether they lie on the well-established red sequence of galaxies. We present the results of this exercise in Figure \ref{fig:color_magnitude}. The red sequence is evident and falls closely to the extrapolated red sequence defined using the colors of normal, low luminosity ($L < L^*$) elliptical galaxies \citep{schombert}. The density ridge of galaxies is nearly indistinguishable from the extrapolated relation. Note that the fractional redshift uncertainties (\S\ref{sec:understanding}), for the dominant fraction of UDGs with accurate redshift estimates, results in absolute magnitudes errors that are smaller than the bin size (0.3 mag) in Figure \ref{fig:color_vs_surface_brightness}.

The presence of UDGs on the extension of the color-magnitude relation suggests that UDGs fall on the stellar mass-metallicity relation (see also \cite{barbosa}). It also argues against certain formation scenarios for the majority of UDGs. For example, tidal dwarfs \citep{hunsberger,duc,Bennet+2018} would be expected to lie off this relation because their stars would come from a more massive parent galaxy. Even though stars from the outskirts of such a massive galaxy would be lower than the characteristic chemical abundance because of metallicity gradients \citep{searle}, it is unlikely that the metallicity of those stars would be on average a match to the extrapolation of the color-magnitude relation. 

These observations can also provide a constraint on
an alternative UDG formation model that posits that UDGs are galaxies that have lost a majority of their initial stellar mass \citep[e.g.,][]{conselice}.  Galaxies with significant stellar mass loss would be expected to have a metallicity above that implied by their current  luminosity. This argument is made more clearly in the bottom panel of Figure \ref{fig:color_magnitude} where we compare the color distribution of UDGs with $-15.5 < {\rm M}_r < -15$ to the position of the extrapolated red sequence (dotted line) and the corresponding mean color of galaxies on that red sequence that lost either 50 or 90\% of their initial stellar mass to match the current M$_r$. Although the scenario where 50\% of the mass is lost is marginally consistent with the observed color distribution, given uncertainties in the extrapolation of the red sequence and photometric calibration, models with larger mass loss become increasingly inconsistent with the observations. We close this discussion by noting that DF 44, one of the best studied UDGs and a close analog in total mass to the Large Magellanic Cloud (LMC) with a mass of $\sim 10^{11}$ M$_\odot$ \citep{vanDokkum+2019,erkal}, has an I-band mass-to-light ratio of $26^{+7}_{-6}$ in solar units \citep{vanDokkum+2019} compared to the LMC's of 4.6 \citep{Kadowaki+2021}). If an LMC-like galaxy was the progenitor of DF 44, then it lost $\sim$ 80\% of its stellar mass.

\subsubsection{Color-Environment Relation}

\begin{figure}
    \centering
    \includegraphics[scale=0.31]{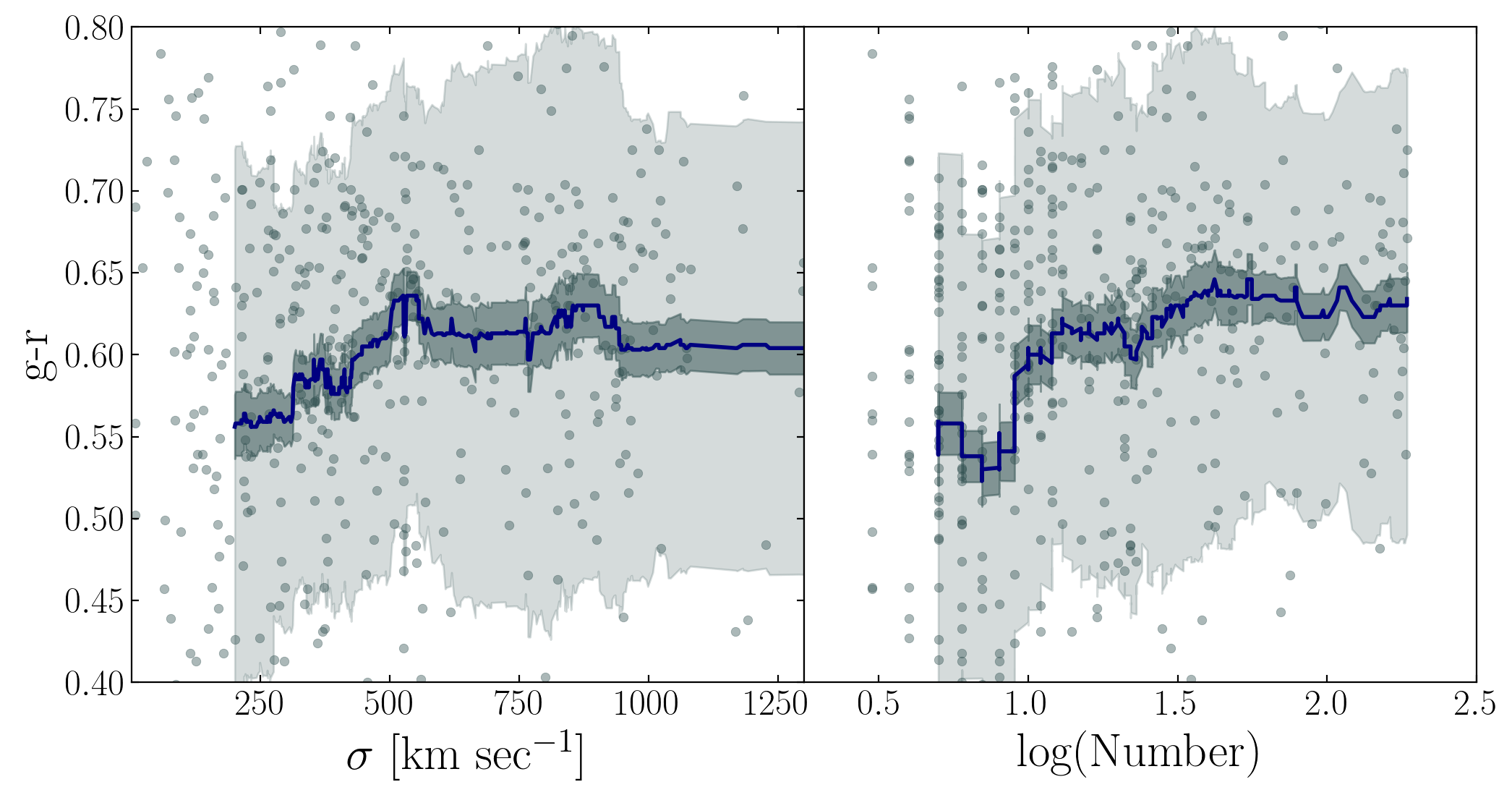}
    \caption{\textit{The dependence of UDG color on environment. In the left panel, we plot UDG color, $g-r$, vs. the line of sight velocity dispersion of the associated overdensity ($\sigma$) for each UDG with an estimated redshift. In the right panel, we plot color vs. the number of galaxies with known redshift in the associated overdensity for each UDG with an estimated redshift. In each panel, the solid line is the rolling mean of 75 galaxies. The lightly shaded area shows the dispersion among measurements in each rolling bin, while the more heavily shaded area shows the dispersion in the mean.}}
    \label{fig:env_dependences}
\end{figure}

We quantify the  environment of each UDG using the standard deviation of the fitted Gaussian to the associated peak of normal galaxies along the line of sight. This measurement is an estimate of the velocity dispersion of the environment. 
In Figure \ref{fig:env_dependences} we plot color, $g-r$ as a function of environmental velocity dispersion. Red, quenched UDGs are found in all environments, while blue, star-forming ones are more highly represented in the lower velocity dispersion environments. This is highlighted by the rolling mean (solid line) which bends at $\sigma \sim 400$ km sec$^{-1}$.
The same qualitative behavior is evident when we categorize environment by the number of normal galaxies found in the associated redshift peak. 
As seen in previous studies \citep{Greco+2018a,Tanoglidis+2021,Kadowaki+2021}, UDG properties track environment, confirming that models of UDGs in isolation will not fully describe them. 


\label{sec:discussion}

\section{Summary}

This paper principally presents a catalog of 5598 ultra-diffuse galaxy (UDG) candidates distributed throughout the southern fraction of the DESI Legacy Imaging Surveys \citep{Dey+2019}, defined as the portion of the survey that used the Blanco 4m telescope and DECam \citep{decam}. By focusing on those candidates that are large on the sky ($r_e >$5.3\arcsec) we aim to limit the number of spurious sources and provide those for which the measured structural and photometric parameters are most accurately determined. The catalog, therefore, is most complete 
for physically large ($r_e > 2.5$ kpc) UDGs lying in the approximate redshift range $1800 \lesssim cz/{\rm km\ s}^{-1} \lesssim 7000$, the lower bound defined by where peculiar velocity uncertainties do not significantly affect the distance estimates and completeness is high and the upper bound by the limits introduced by our angular selection criterion. 

Because the definition of UDG incorporates a physical size criterion, we proceed to develop a methodology for estimating distances based on an extension of the way the original UDG samples were constructed, using distance-by-association. In distance-by-association those UDG candidates projected near evident overdensities \citep[e.g., the Coma cluster;][]{vanDokkum+2015} are placed at the distance of the overdensity. We extend the method to lower amplitude, uncatalogued overdensities and obtain preliminary estimated redshifts for 1050 candidates. We find that the redshifts are accurate for between 70 and 90\% of the sources. Of those with estimated redshifts, 514 satisfy the $r_e \ge 1.5$ kpc criterion for UDGs and lie at $cz > 1800$ km sec$^{-1}$. We consider these preliminary redshift estimates because as the sample of UDG candidates with spectroscopic redshifts increases we will be able to train the method further and refine the sample of estimated redshifts that we consider to be accurate.

Finally, we present a sampling of results drawn from the catalog to illustrate its uses. We present results that are distance independent and dependent. In the former we revisit results from Paper II, and in the latter we establish that the red sequence of UDGs follows closely the extrapolation of the red sequence relation for bright ellipticals and that the environment-color relation is at least qualitatively similar to that of high surface brightness galaxies. Both of these results challenge some of the models proposed for UDG evolution, and more detailed examination of the catalog will surely provide further constraints. Modelers now have additional empirical results to target. 

\begin{acknowledgments}

DZ, RD, JK, and HZ acknowledge financial support from NSF AST-1713841 and AST-2006785. AD's research is supported by NOIRLab, the John Simon Guggenheim Foundation, and the Institute of Theory and Computation at the Harvard-Smithsonian Center for Astrophysics. KS acknowledges funding from the Natural Sciences and Engineering Research Council of Canada (NSERC). An allocation of computer time from the UA Research Computing High Performance Computing (HPC) at the University of Arizona and the prompt assistance of the associated computer support group is gratefully acknowledged.
 
This research has made use of the NASA/IPAC Extragalactic Database (NED), which is operated by the Jet Propulsion Laboratory, California Institute of Technology, under contract with NASA. 

This research depends directly on images from the Dark Energy Camera Legacy Survey (DECaLS; Proposal ID 2014B-0404; PIs: David Schlegel and Arjun Dey). Full acknowledgment at \url{https://www.legacysurvey.org/acknowledgment/}.

\end{acknowledgments}

\facilities{Blanco/DECam}

\software{
Astropy              \citep{astropy1, astropy2},
astroquery           \citep{astroquery},
GALFIT               \citep{peng},
keras                \citep{keras},
lmfit                \citep{newville},
Matplotlib           \citep{matplotlib},
NumPy                \citep{numpy},
pandas               \citep{pandas},
sep                  \citep{sep},
Source Extractor     \citep{bertin},
SciPy                \citep{scipy1, scipy2},
Scikit-learn         \citep{scikit-learn}
SWarp                \citep{Swarp}}
TensorFlow          \citep{tensorflow}

\bibliography{refs.bib}
\bibliographystyle{aasjournal}

\end{document}